%% file: main.tex
\documentclass{article}
\usepackage[utf8]{inputenc}
\usepackage{graphicx}
\usepackage[portrait, margin=0.8in]{geometry}
\usepackage{hyperref}
\usepackage{authblk}
\usepackage{todonotes}
\usepackage{algpseudocode}
\usetikzlibrary{quantikz}
\usetikzlibrary{external}
\usepackage{booktabs,multirow}
\usepackage{adjustbox}
\usepackage{tikz}
\usepackage{subcaption}
\usepackage{braket}
\usepackage{amsopn}
\usepackage{amssymb,amsfonts}
\usepackage{caption}
\usepackage{algorithm,algorithmicx,algcompatible}
\usepackage{braket}
\usepackage{amsmath,amsthm,amssymb}
\usepackage[sorting = none, backend = bibtex, style=numeric-comp]{biblatex}

\usepackage{tikz}
\usetikzlibrary{shapes.geometric, arrows}

\tikzstyle{startstop} = [rectangle, rounded corners, 
minimum width=3cm, 
minimum height=1cm,
text centered, 
draw=black, 
fill=red!30]

\tikzstyle{io} = [trapezium, 
trapezium stretches=true, 
trapezium left angle=70, 
trapezium right angle=110, 
minimum width=3cm, 
minimum height=1cm, text centered, 
draw=black, fill=blue!30]

\tikzstyle{process} = [rectangle, 
minimum width=3cm, 
minimum height=1cm, 
text centered, 
text width=3cm, 
draw=black, 
fill=orange!30]

\tikzstyle{decision} = [diamond, 
minimum width=3cm, 
minimum height=1cm, 
text centered, 
draw=black, 
fill=green!30]
\tikzstyle{arrow} = [thick,->,>=stealth]

\title{Increasing the Measured Effective Quantum Volume with Zero Noise Extrapolation}

\author[1]{Elijah Pelofske\thanks{Email: epelofske@lanl.gov}}
\author[2]{Vincent Russo\thanks{Email: vincent@unitary.fund}}
\author[2,3]{Ryan LaRose}
\author[2,5]{Andrea Mari}
\author[2]{Dan Strano}
\author[1]{Andreas Bärtschi}
\author[1]{Stephan Eidenbenz}
\author[2,4]{William J. Zeng}

\affil[1]{CCS-3 Information Sciences, Los Alamos National Laboratory}
\affil[2]{Unitary Fund}
\affil[3]{Institute of Physics, \'{E}cole Polytechnique F\'{e}d\'{e}rale de Lausanne (EPFL), CH-1015 Lausanne, Switzerland}
\affil[4]{Goldman Sachs \& Co, New York, NY}
\affil[5]{Physics Division, School of Science and Technology, Universit\`a di Camerino, Italy}

\date{\vspace{-6ex}}

\begin{document}
\maketitle

\input{abstract}
\input{text}

\end{document}

%% file: abstract.tex
\begin{abstract}
Quantum Volume is a full-stack benchmark for near-term quantum computers. It quantifies the largest size of a square circuit which can be executed on the target device with reasonable fidelity. Error mitigation is a set of techniques intended to remove the effects of noise present in the computation of noisy quantum computers when computing an expectation value of interest. Effective quantum volume is a proposed metric that applies error mitigation to the quantum volume protocol in order to evaluate the effectiveness not only of the target device but also of the error mitigation algorithm. Digital Zero-Noise Extrapolation (ZNE) is an error mitigation technique that estimates the noiseless expectation value using circuit folding to amplify errors by known scale factors and then extrapolating computed expectation values to the zero-noise limit. Here we demonstrate that ZNE, with global and local unitary folding with fractional scale factors, in conjunction with dynamical decoupling, can increase the effective quantum volume over the vendor-measured quantum volume. Specifically, we measure the effective quantum volume of four IBM Quantum superconducting processor units, obtaining values that are larger than the vendor-measured quantum volume on each device. This is the first such increase reported.
\end{abstract}

%% file: text.tex
\section{Introduction}
\label{section:introduction}
Quantum Volume (QV)~\cite{Quantum_Volume, moll2018quantum, QV_64_IBMQ, QV_in_practice, Baldwin2022reexaminingquantum} is a proposed full-stack metric for evaluating the capability of error-prone quantum computers in the NISQ era~\cite{Preskill_2018}. The intention of the Quantum Volume protocol is to measure the largest \emph{square} circuit acting on $n$ qubits with depth $d=n$ which can be computed with reasonably high fidelity on a target Quantum Processing Unit (QPU)~\footnote{Typically, a quantum computer passing the QV protocol for $n$ qubits is said to a have a quantum volume of $2^n$. Here, we will simply use the notation of $n$ instead of $2^n$.}. A QV circuit is composed of alternating layers of random qubit index permutations and random SU(4) two-qubit operations. The logical diagram of a QV circuit is given in Figure~\ref{fig:logical_QV_circuit}. Each of the $d$ layers of two-qubit SU(4) gate operations act on $n' := \lfloor \frac{n}{2} \rfloor $ pairs of qubits, meaning that if $n$ is odd then one qubit will be idle in each layer. The Quantum Volume protocol is based on the \emph{heavy output generation problem}~\cite{10.5555/3135595.3135617}, where a single QV circuit $U$ has an ideal bitstring output distribution given in eq.~\eqref{equation:U_prob}

\begin{equation}
    p_U(x) = |\braket{x|U|0}|^2.
    \label{equation:U_prob}
\end{equation}

Given the ideal distribution $p_U(x)$ for a single QV circuit $U$, each probability for all $2^n$ possible states is sorted $p_1 \leq p_2 \dots \leq p_{2^n}$. Eq.~\eqref{equation:H_U} defines the set of heavy bitstrings $H_U$, given the median of all of the probabilities $p_{\mathrm{median}}$ and all of the classically computed bitstring probabilities 
$p_U(x)$, 
\begin{equation}
    H_U = \{ x \in \{0, 1\}^n : p_U(x) > p_{\mathrm{median}}\}.
    \label{equation:H_U}
\end{equation}
Given the set $H_U$, the heavy output probability (HOP) is the probability of measuring a bitstring which belongs to $H_U$. Empirically, given the measured bitstrings obtained by executing a Quantum Volume circuit, the HOP can be estimated by taking the proportion of the measurements which are in the set $H_U$, out of the total number of measurements that were made.

As it was originally proposed, the QV protocol~\cite{Quantum_Volume, moll2018quantum} specifies that some reasonably large number of samples (e.g. at least $10$) is measured from at least $100$ randomly generated QV circuits that are compiled and executed on a target quantum computer which we wish to measure the computational capabilities of. Therefore, for each QV circuit, we can estimate the associated HOP and, by averaging over all QV circuits, we obtain an estimate of the mean HOP. For a noiselss quantum computation the mean HOP approaches $\frac{1 + \ln(2)}{2} \approx 0.85$~\cite{10.5555/3135595.3135617, Quantum_Volume} in the limit of large $n$. The QV protocol~\cite{Quantum_Volume} passes for size $n$ if a quantum computer is able to reliably sample an ensemble of random QV circuits at a mean HOP greater than $\frac{2}{3}$. If the computation loses all coherence, and the circuit is effectively randomly sampling bits, then the mean HOP converges to $0.5$.

\newcommand{\sugate}{\gate[2]{\mathrm{SU}(4)}}
\newcommand{\pigate}[2]{\gate[#1]{\pi_{#2}}}
\begin{figure}[!t]
    \centering
	\begin{adjustbox}{width=0.6\linewidth}
		\begin{quantikz}[row sep={24pt,between origins},execute at end picture={
					\node at ($(\tikzcdmatrixname-7-2)+(28pt,-15pt)$) {\textcolor{red}{1}};	
					\node at ($(\tikzcdmatrixname-7-4)+(28pt,-15pt)$) {\textcolor{red}{2}};	
					\node at ($(\tikzcdmatrixname-7-10)+(28pt,-15pt)$) {\textcolor{red}{$d$}};	
			}]
			\lstick{$\ket{0}$}\slice{}	& \pigate{7}{1}	& \sugate\slice{}	& \pigate{7}{2}	& \sugate\slice{}	& \qw	& \ldots	& 	& \qw\slice{}	& \pigate{7}{d}	& \sugate\slice{}	& \meter{} 	\\
			\lstick{$\ket{0}$}		& \qw		& \qw			& \qw		& \qw			& \qw	& \ldots	& 	& \qw		& \qw		& \qw			& \meter{} 	\\
			\lstick{$\ket{0}$}		& \qw		& \sugate		& \qw		& \sugate		& \qw	& \ldots	& 	& \qw		& \qw		& \sugate		& \meter{} 	\\
			\lstick{$\ket{0}$}		& \qw		& \qw			& \qw		& \qw			& \qw	& \ldots	& 	& \qw		& \qw		& \qw			& \meter{} 	\\
			\lstick{$\ket{0}$}		& \qw		& \sugate		& \qw		& \sugate		& \qw	& \ldots	& 	& \qw		& \qw		& \sugate		& \meter{} 	\\
			\lstick{$\ket{0}$}		& \qw		& \qw			& \qw		& \qw			& \qw	& \ldots	& 	& \qw		& \qw		& \qw			& \meter{} 	\\
			\lstick{$\ket{0}$}		& \qw		& \qw			& \qw		& \qw			& \qw	& \ldots	& 	& \qw		& \qw		& \qw			& \meter{} 	
		\end{quantikz}
	\end{adjustbox}
	\caption{Circuit drawing for a logical QV circuit acting on $n=7$ qubits, where each layer $d=1,\ldots,d=7$ consists of a random permutation of the qubit labels 
	followed by $n' = 3 =\lfloor \tfrac{7}{2} \rfloor$ random SU(4) gates acting on consecutive pairs of (the permuted) qubits. This figure was realized using Quantikz~\cite{quantikz}.}
    \label{fig:logical_QV_circuit}
\end{figure}
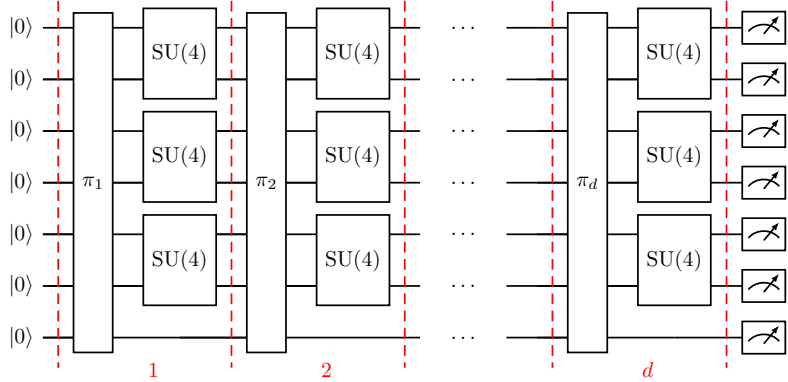

\begin{table*}[!t]
\begin{center}
\begin{tabular}{ |p{2.3cm}|p{2.6cm}|p{3.0cm}|p{2.9cm}|p{2.9cm}| }
 \hline
 QPU name & IBM Quantum processor type & Number of qubits & Vendor-measured Quantum Volume & Measured (ZNE) effective  quantum volume \\ 
 \hline
 \hline
 \texttt{ibmq\_toronto} & Falcon r4 & 27 & 5 & 6 \\ 
 \hline
 \texttt{ibm\_geneva} & Falcon r8 & 27 & 5 & 6 \\ 
 \hline
 \texttt{ibm\_auckland} & Falcon r5.11 & 27 & 6 & 7 \\ 
 \hline
 \texttt{ibm\_hanoi} & Falcon r5.11 & 27 & 6 & 7 \\ 
 \hline
\end{tabular}
\end{center}
\caption{IBM Quantum superconducting processor summary from the experiments presented in this work. The basis gates for all four of these devices are \texttt{CX, ID, RZ, SX, X}. Quantum Volume and Effective Quantum Volume numbers are reported as $n$ instead of $2^n$. The Vendor-measured Quantum Volume values were recorded from the vendor at the time these experiments were performed. }
\label{table:IBMQ_hardware_summary}
\end{table*}

Quantum error mitigation, alongside quantum error suppression, is a family of techniques that can be utilized to remove or suppress errors in a computation performed on a NISQ computer~\cite{https://doi.org/10.48550/arxiv.2210.00921, Kandala_2019, https://doi.org/10.48550/arxiv.2301.02690, PhysRevX.8.031027, PRXQuantum.2.040330}. Quantum error mitigation is thought of as an intermediate step in the development of quantum computers, eventually leading to the development of fully realized quantum error correction~\cite{https://doi.org/10.48550/arxiv.2208.01863, https://doi.org/10.48550/arxiv.2207.06431}. The trade off that comes from using quantum error mitigation is that in general there is significant computational cost of applying quantum error mitigation, namely in terms of additional compute time and measurements that must be obtained. In particular, the number of samples required to measure a certain accuracy of an observable on a noisy quantum computer scales exponentially with circuit depth \cite{Takagi_2022, quek2023exponentially}. 

One of the properties of Quantum Volume that makes it very relevant for evaluating quantum computing performance is that it involves the full stack of software and hardware---including what parts of the hardware are used, and what compilation is used~\cite{QV_in_practice}. Effective quantum volume~\cite{effective_QV} is a proposed variant of the Quantum Volume metric which uses error mitigation strategies in conjunction with Quantum Volume in order to evaluate the quality of quantum computations that can be performed on modern quantum computers with error mitigation. Zero Noise Extrapolation (ZNE)~\cite{PhysRevA.102.012426, PhysRevX.7.021050, Kandala_2019, Giurgica_Tiron_2020, Schultz_2022, Temme_2017, LaRose2022mitiqsoftware, PhysRevA.105.042406, Kim_2023, tran2023locality, Kim2023} is a general-purpose error mitigation algorithm that uses expectation values computed at different noise levels so as to extrapolate the computation to its zero noise limit. The different noise levels can be indirectly created by circuit folding operations that leave the circuit invariant in its theoretical expectation value, but purposely increase the number of instructions (gates) which means there will be more errors within the computation on the NISQ computer, such that the amount of noise within the computation will have been scaled by a pre-determined scale factor $\lambda$. The effect of ZNE when applied to a Quantum Volume circuit is to effectively increase the error-mitigated HOP, obtaining a result closer to the ideal limit of $\frac{1 + \ln(2)}{2}$. Informally, the Zero Noise Extrapolation Quantum Volume benchmark, which we refer to as the \emph{effective quantum volume}~\cite{effective_QV}, is defined such that a quantum computer that passes the test for square dense random circuits of depth $n$ and acting on $n$ qubits will be able to compute reasonably good expectation values using ZNE for general quantum circuits acting on $n$ qubits and whose depth is at most $n$. Specifically, for a NISQ computer if the zero noise extrapolated Heavy Output Probability expectation values pass (with high confidence) the $\frac{2}{3}$ threshold, as defined in the original quantum volume benchmark, for $n$ by $n$ square quantum volume circuits, then we say that the device has an \emph{effective quantum volume} of $n$. We expect, in general, the \emph{effective quantum volume} number to be greater than the original quantum volume benchmark number for NISQ devices. The effective quantum volume protocol~\cite{effective_QV} can be scaled to large system sizes; however there are two primary limitations of its scalability. The first comes from the original quantum volume protocol, which is that we need to classically compute the full statevector of the random quantum circuits -- and this becomes prohibitive even for HPC clusters at qubit counts greater than $\sim 32$. The second limitation comes from the inherent noise level in the quantum computer being benchmarked; this error rate is the limiting factor for arbitrarily increasing, and passing the test for, the \emph{effective quantum volume} size.

In this study we experimentally demonstrate increased effective quantum volume compared to the vendor-benchmarked QV values, using ZNE, on four IBM Quantum devices.  
The results are summarized in Table~\ref{table:IBMQ_hardware_summary}\footnote{Note that since these experiments were executed, both \texttt{ibm\_geneva} and \texttt{ibmq\_toronto} have been decommissioned}.

The dataset produced from this study is publicly available at~\cite{pelofske_2024_12629044}.

\section{Methods}
\label{section:methods}

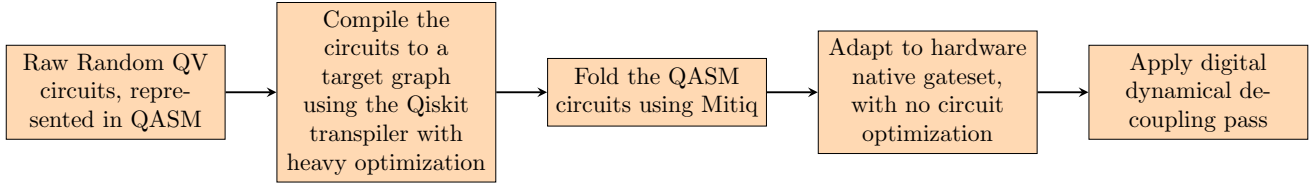
\begin{figure}[!htpb]
    \centering
    \noindent\resizebox{\textwidth}{!}{
    \begin{tikzpicture}[node distance=2cm]
        \node (1a) [process] {Raw Random QV circuits, represented in QASM};
        \node (1b) [process, right of=1a, xshift=2cm] {Compile the circuits to a target graph using the Qiskit transpiler with heavy optimization};
        \node (1c) [process, right of=1b, xshift=2cm] {Fold the QASM circuits using Mitiq};
        \node (1d) [process, right of=1c, xshift=2cm] {Adapt to hardware native
        gateset, with no circuit optimization};
        \node (1e) [process, right of=1d, xshift=2cm] {Apply digital dynamical decoupling pass};

        \draw [arrow] (1a) -- (1b);
        \draw [arrow] (1b) -- (1c);
        \draw [arrow] (1c) -- (1d);
        \draw [arrow] (1d) -- (1e);
    \end{tikzpicture}
    }
    \caption{High-level workflow of the methodology for estimating the effective quantum volume with zero-noise
extrapolation. First, the raw uncomputed QV circuits are initially defined with arbitrary connectivity and basis gates \texttt{U3} and \texttt{CX}. Second, the Qiskit transpiler is run on the target hardware subgraph with \texttt{optimization\_level=3}. Third, the QASM
circuits are folded via either global or local folding with Mitiq. Fourth, the native
gateset is adapted to \texttt{RZ}, \texttt{CX}, \texttt{SX}, and \texttt{X} using the Qiskit transpiler with \texttt{optimization\_level=0}. Fifth, an optional step, X-X digital dynamical decoupling is applied.
}
    \label{fig:workflow_diagram}
\end{figure}

The QV ZNE protocol steps that we use for this study are outlined at a high level in Figure~\ref{fig:workflow_diagram}. 

\subsection{Generation of quantum volume circuits}

\begin{figure}[!ht]
    \centering
    \includegraphics[width=0.23\textwidth]{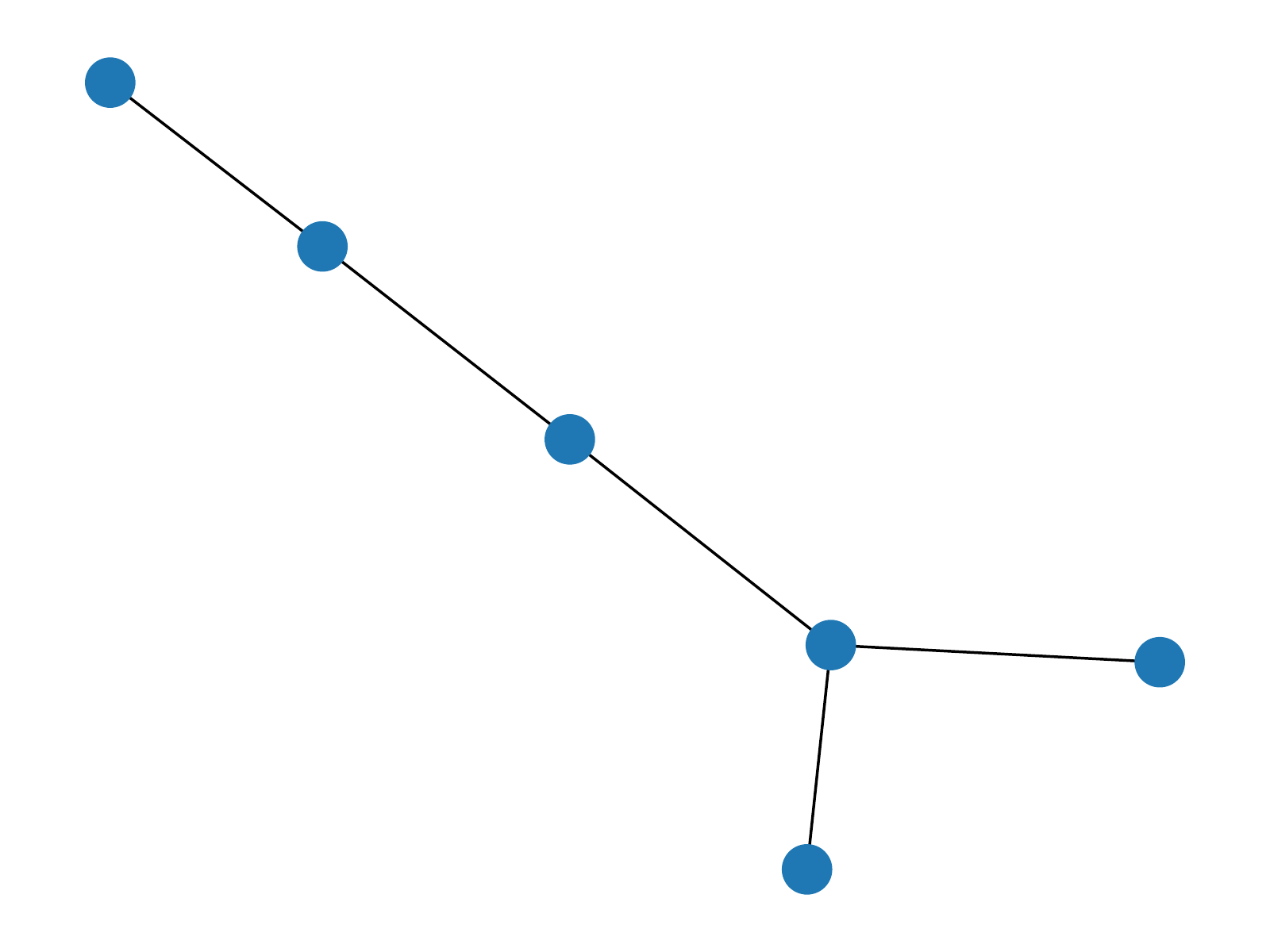}
    \includegraphics[width=0.23\textwidth]{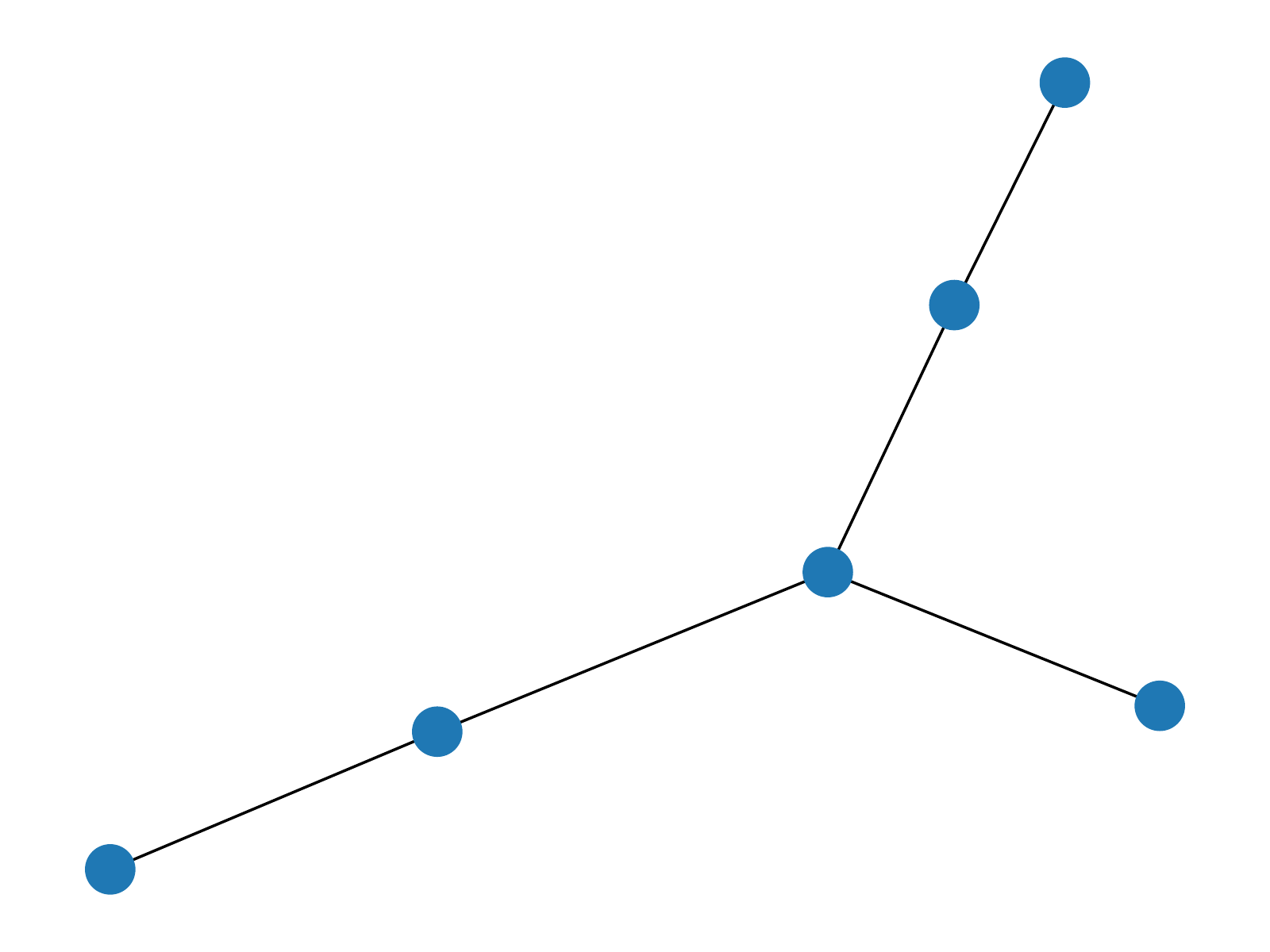}
    \includegraphics[width=0.23\textwidth]{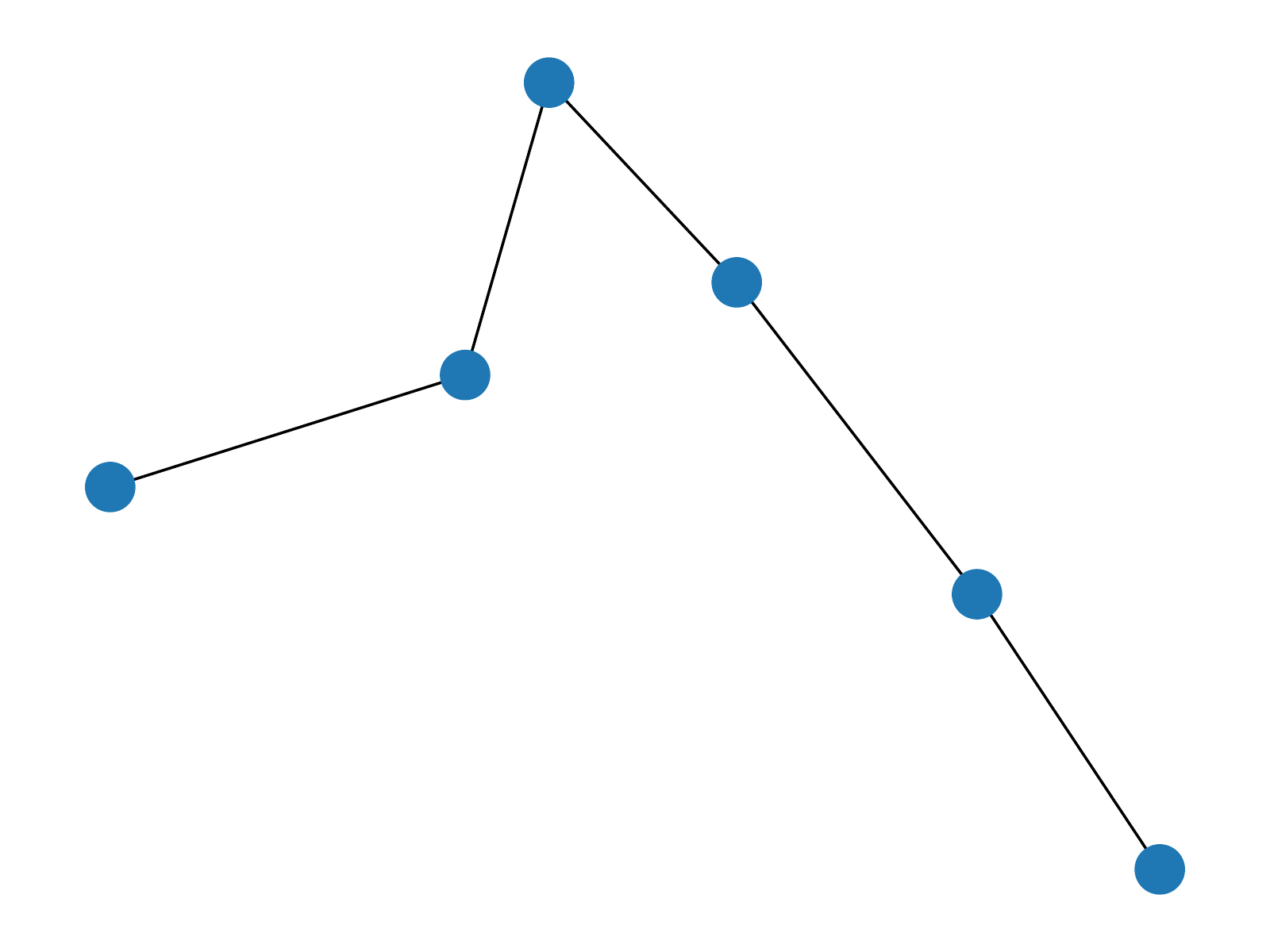}
    \caption{All subgraph isomorphisms for $n=6$ on the heavy-hex connectivity.}
    \label{fig:isomprhisms_ibmq_montreal_n6}
\end{figure}

\begin{figure}[!ht]
    \centering
    \includegraphics[width=0.19\textwidth]{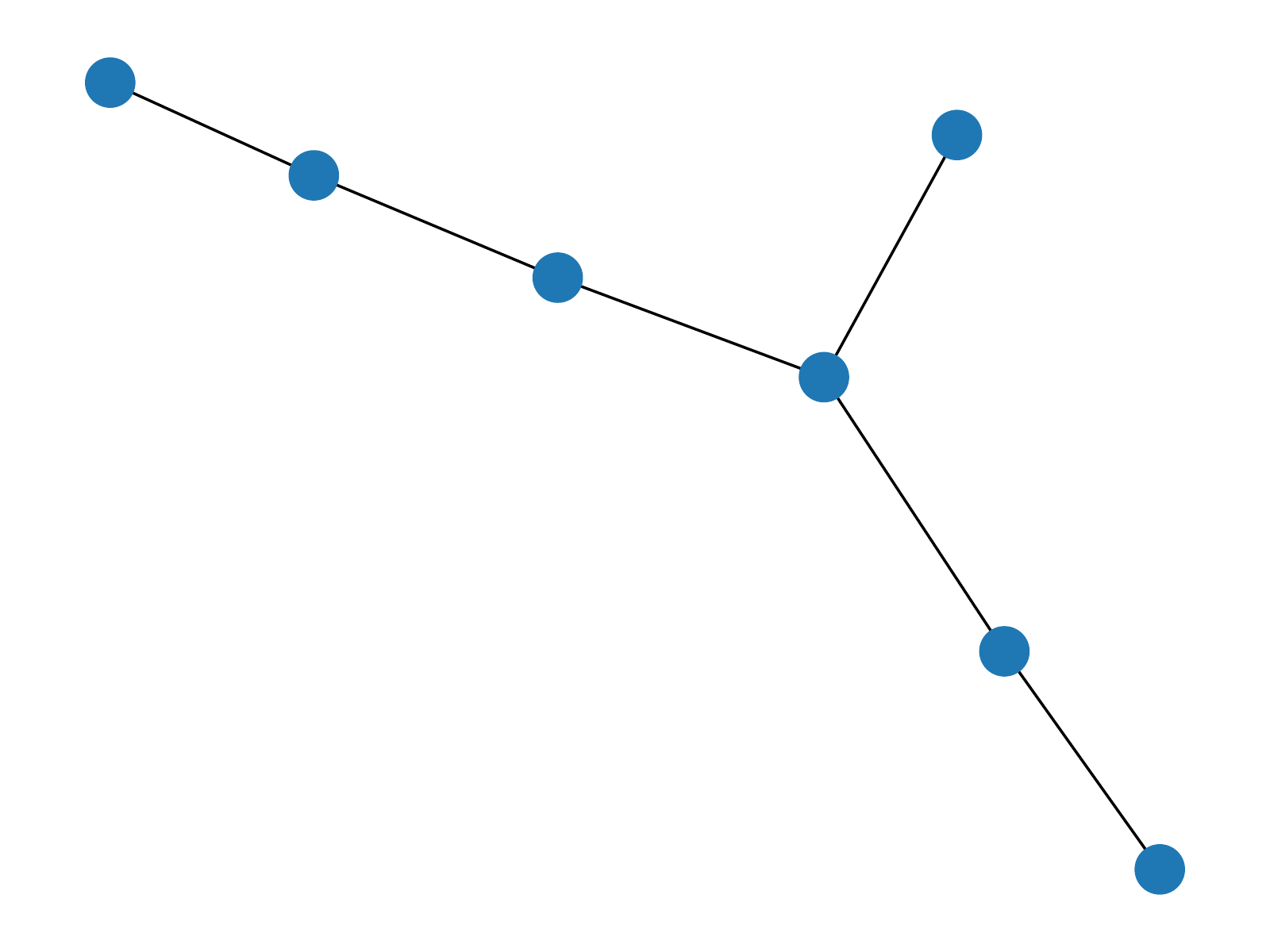}
    \includegraphics[width=0.19\textwidth]{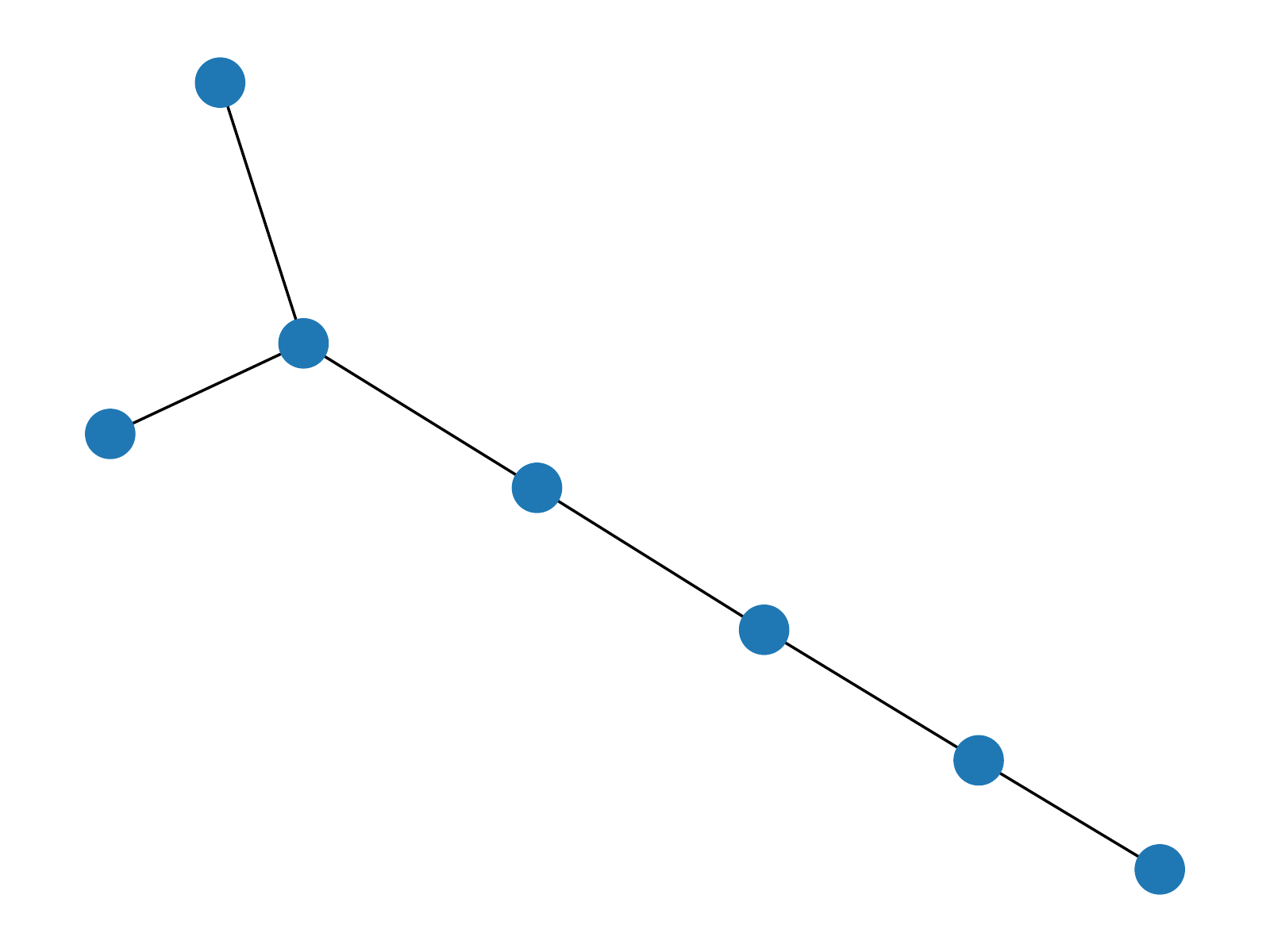}
    \includegraphics[width=0.19\textwidth]{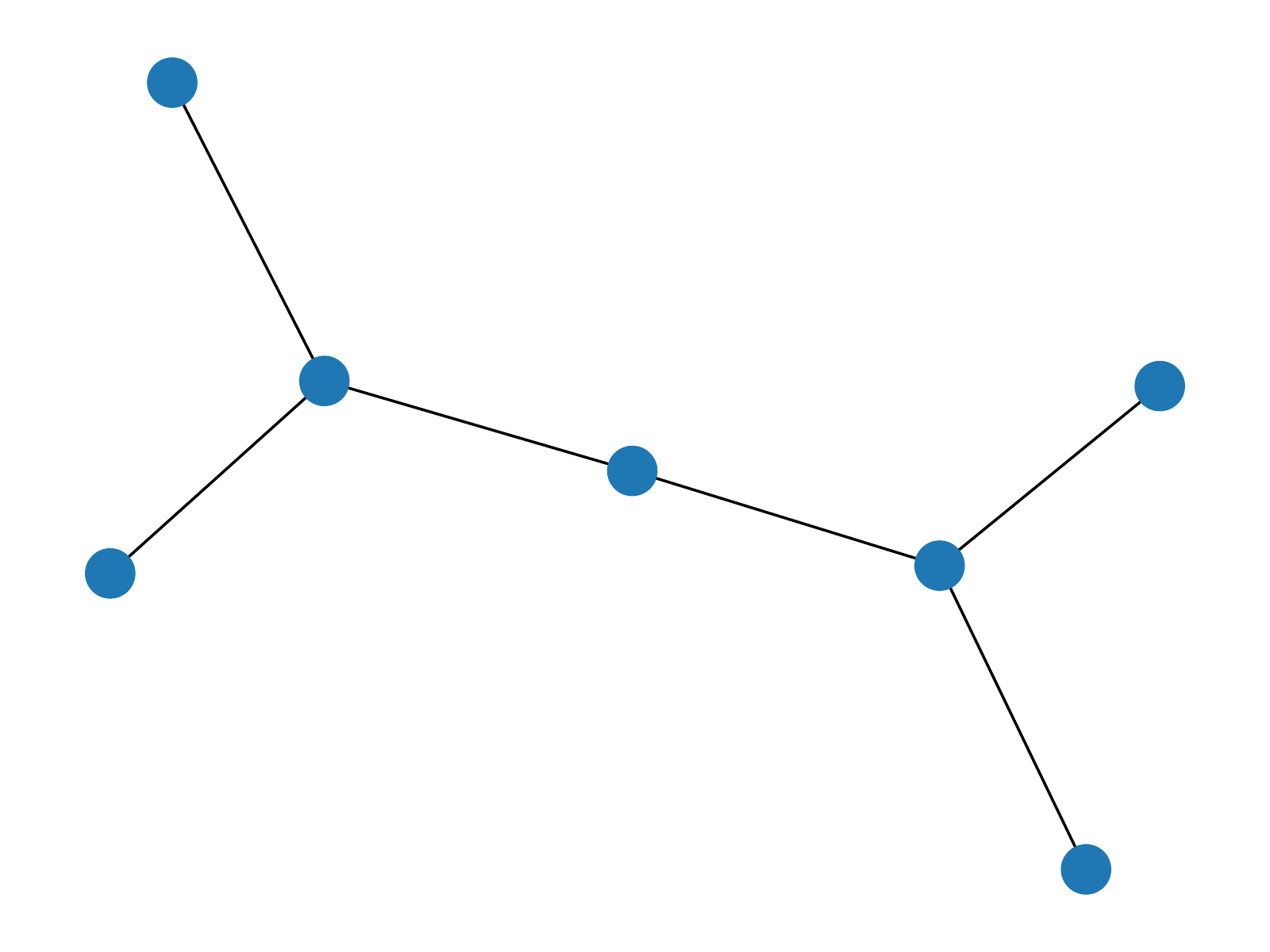}
    \includegraphics[width=0.19\textwidth]{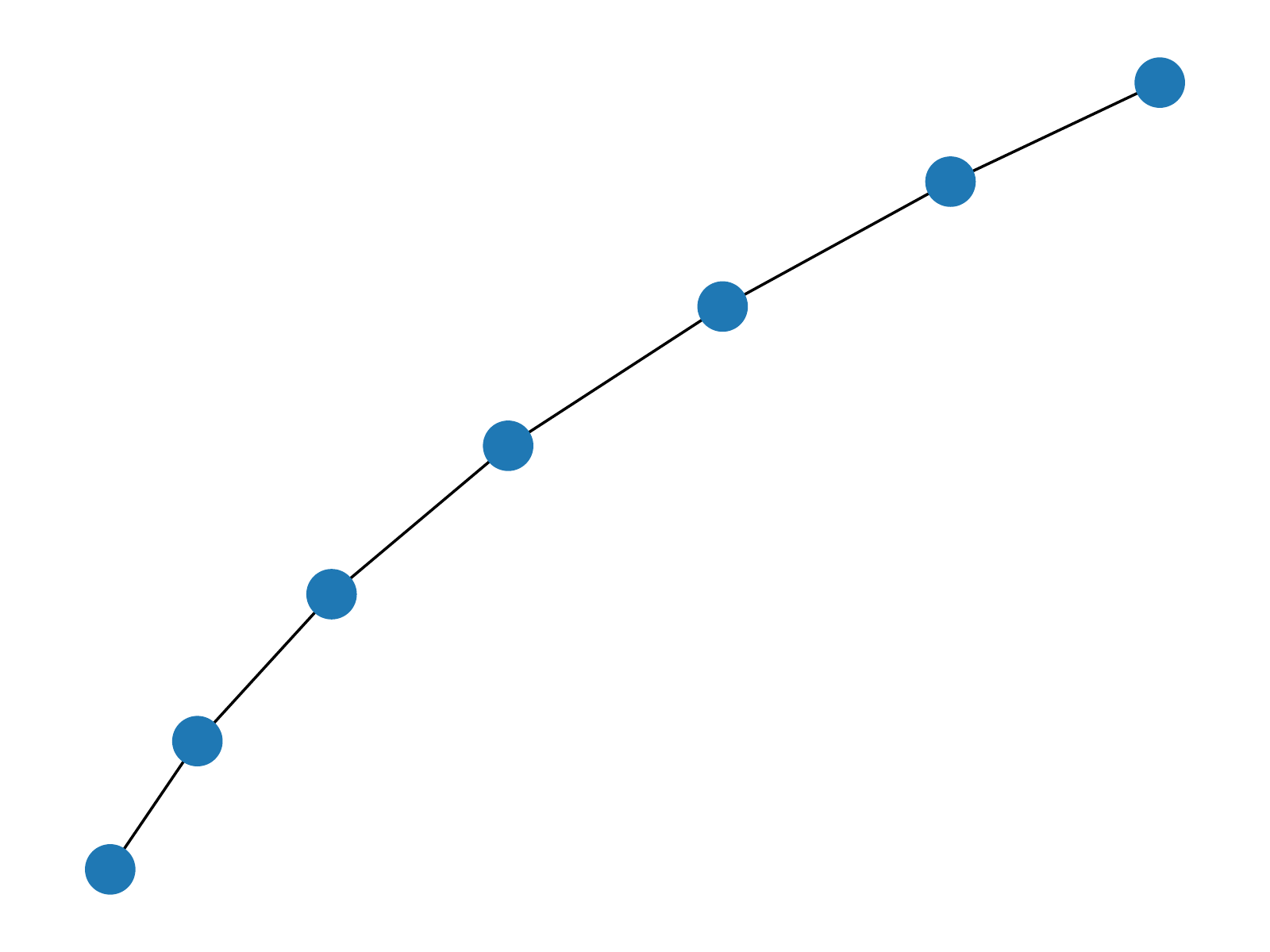}
    \includegraphics[width=0.19\textwidth]{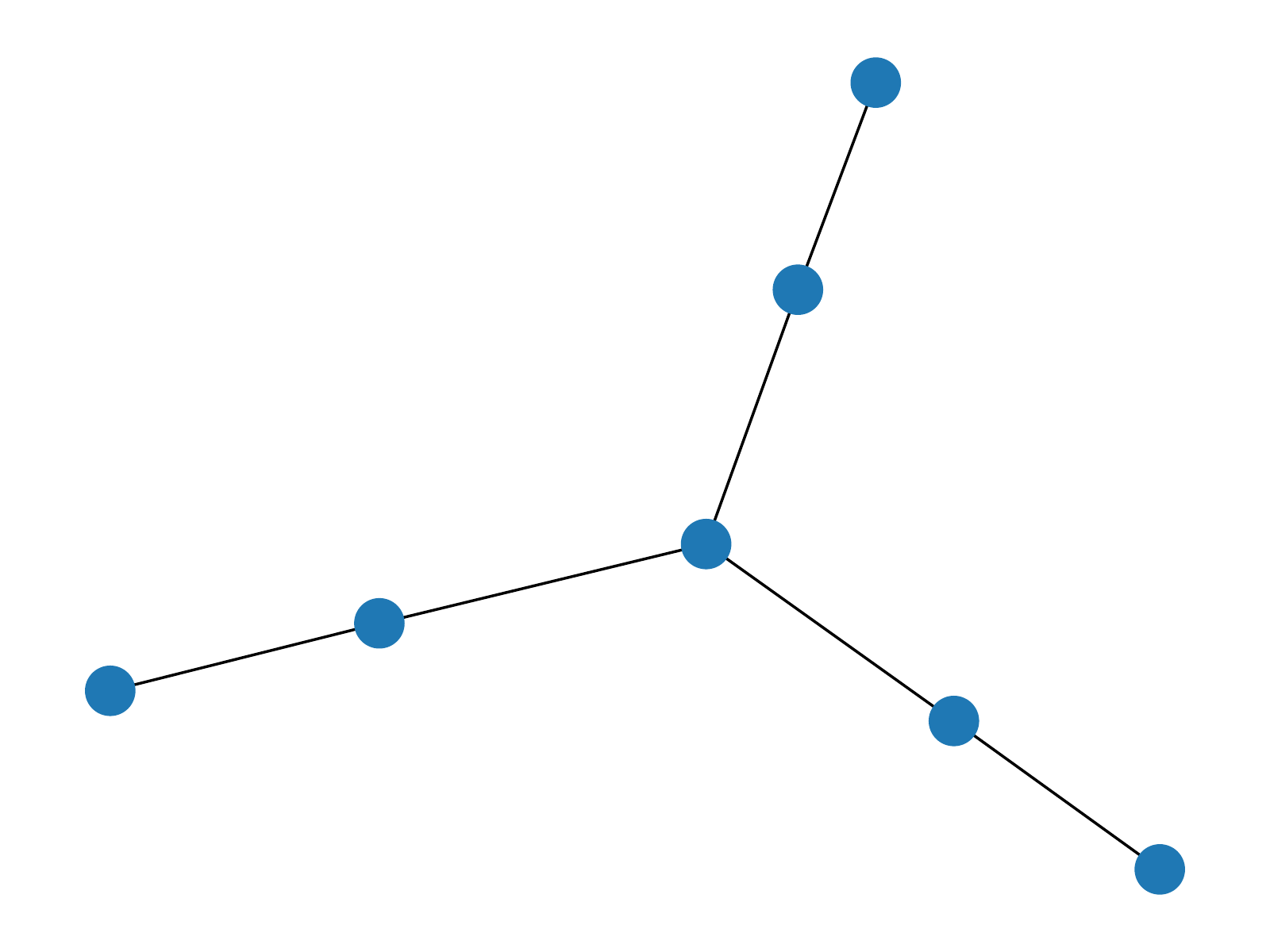}
    \caption{All subgraph isomorphisms for $n=7$ on heavy hex connectivity.}
    \label{fig:isomprhisms_ibmq_montreal_n7}
\end{figure}

The first step in executing the ZNE QV benchmark protocol is to construct raw, random, Quantum Volume circuits of the form shown in Figure~\ref{fig:logical_QV_circuit}. In this case, the raw QV circuits are generated using Qiskit~\cite{Qiskit} and are defined with arbitrary connectivity and native gates of \texttt{CX} (e.g. CNOT) and \texttt{u3} (e.g. arbitrary single qubit rotation). Next, the raw QV circuits are transpiled using a single call to the Qiskit~\cite{Qiskit} transpiler (\texttt{optimization\_level} set to the maximum of $3$) to target a specific device, thereby requiring a specific qubit connectivity and a specific native gateset~\cite{QV_in_practice}. In this case, the native gateset is \texttt{CX, RZ, X, SX}. The target connectivity is a subgraph of a heavy-hex~\cite{PhysRevX.10.011022} hardware graph. In order to reduce compilation times, each of the possible subgraph isomorphisms for a given size $n$ are compiled to (for all QV circuits). These possible isomorphic subgraphs for $n=6$ and $n=7$ are shown in Figures~\ref{fig:isomprhisms_ibmq_montreal_n6} and~\ref{fig:isomprhisms_ibmq_montreal_n7}. This compilation method allows us to compile the QV circuits to a small number of different connectivities, and then map those compiled circuits to any part of the quantum device chip without the need to re-compile all of the QV circuits. This reduces the total (classical) compute time required for executing these experiments. Furthermore, by choosing specific layouts to compile the circuits to, this allows us to control specifically which parts of the hardware graph are used, and then enumerate over different parts of the chip. Lastly, since the compiled circuits for a single sub-graph isomorphism are the same no matter which isomorphic subgraph they are mapped to on the device, this allows the results to be directly comparable (whereas, the stochasticity of the quantum circuit compiler can result in slightly different compiled circuits if the compilation is re-executed for every target subgraph). When executing the circuits on the IBM Quantum devices, we enumerated over multiple hardware subgraphs, similar to ref.~\cite{QV_in_practice}, but did not enumerate over \emph{every possible} hardware subgraph as in ref.~\cite{QV_in_practice} because the ZNE protocol adds significant time and computation overhead to the already time-intensive Quantum Volume protocol. However, we did enumerate over as many hardware subgraphs as could be executed, given the constraints of queue times and server-side device errors.

\subsection{Noise scaling by unitary folding}

Once each circuit was compiled to the fixed part of the hardware, the circuits were then {\it folded} (i.e.\ compiled with more gates) using the Mitiq software package~\cite{LaRose2022mitiqsoftware} so as to scale up the level of noise. Specifically, we used two different techniques known as global folding and random local folding, which are both designed to effectively increase the error rate when running the circuits  on a noisy quantum processor.~\cite{PhysRevA.102.012426, Temme_2017, Giurgica_Tiron_2020, PhysRevA.102.012426, Kandala_2019, PhysRevX.7.021050}.

Let $U=L_d L_{d-1} \dots L_1$ be the a decomposition of the full circuit as a sequence of $d$ layers and let $W=L_d L_{d-1} ... L_{d -k-1}$ be the fraction of the full circuit corresponding to the last $k$ layers. Global folding transforms the input circuit $U$ as follows
\footnote{Here and everywhere in this work, we assume $\lambda \le 3$. More details about how to extend global and local folding for $\lambda>3$ can be found in Refs.~\cite{Giurgica_Tiron_2020, LaRose2022mitiqsoftware}. }
\begin{equation}\label{eq:global_folding}
U \rightarrow  U W^{\dag} W,
\end{equation}
where $k$ depends on the noise scale factor $\lambda$. More precisely, $k$ is chosen as the integer part of $d(\lambda -1)/2$
, such that the new depth $d'=d+2k$ is approximately scaled as  $\lambda d$.
For the local folding technique, the transformation is similar but, instead of being applied at the circuit level, it is applied at the gate level as follows:
\begin{equation}\label{eq:local_folding}
G \rightarrow  G G^\dag G,
\end{equation}
where $G$ represents an individual gate of the input circuit $U$. Let $t$ be the number of noisy gates in the input circuit $U$ and let $k$ be the integer part of $t(\lambda - 1)/2$. In random local folding, the transformation in Eq.\ \eqref{eq:local_folding} is applied to $k$ noisy gates sampled at random without replacement from the full set of the $t$ noisy gates in the circuit, such that the new circuit has $\approx \lambda t$ noisy gates. More precisely, in this work we assume that the main source of errors is due to CNOT gates and, therefore, we apply local folding by only sampling from the set of $t$ CNOT gates that are present in each quantum volume circuit.
Moreover, to apply local folding more uniformly along the circuit, the procedure was repeated $10$ times for each noise scale factor $\lambda$ (each time sampling a new random subset of $k$ CNOT gates), and the measured expectation values averaged to a single noise-scaled value.

For all the experiments reported in this work, the chosen noise scale factors are $\lambda = \{1, 1.2, 1.5, 1.8, 2\}$ (or a subset of them). These scale factors are quite small compared to those used in the quantum volume experiments of  Ref.~\cite{effective_QV}. Using small scale factors is a deliberate choice since in this study we consider larger circuits for which large noise scale factors are not appropriate. Indeed the application of strong noise scaling (e.g. with $\lambda \ge 3$) would amplify errors too much when applied on a base circuit which is already very noisy.

For storing the circuits at this stage, OpenQASM is used~\cite{qasm}. After the noise scaling process, the circuits are again transpiled to map them to the correct gateset, but no optimization is applied such that the intended effect of unitary folding is not cancelled by circuit optimization passes.

\begin{figure}[t!]
    \centering
    \includegraphics[width=0.95\textwidth]{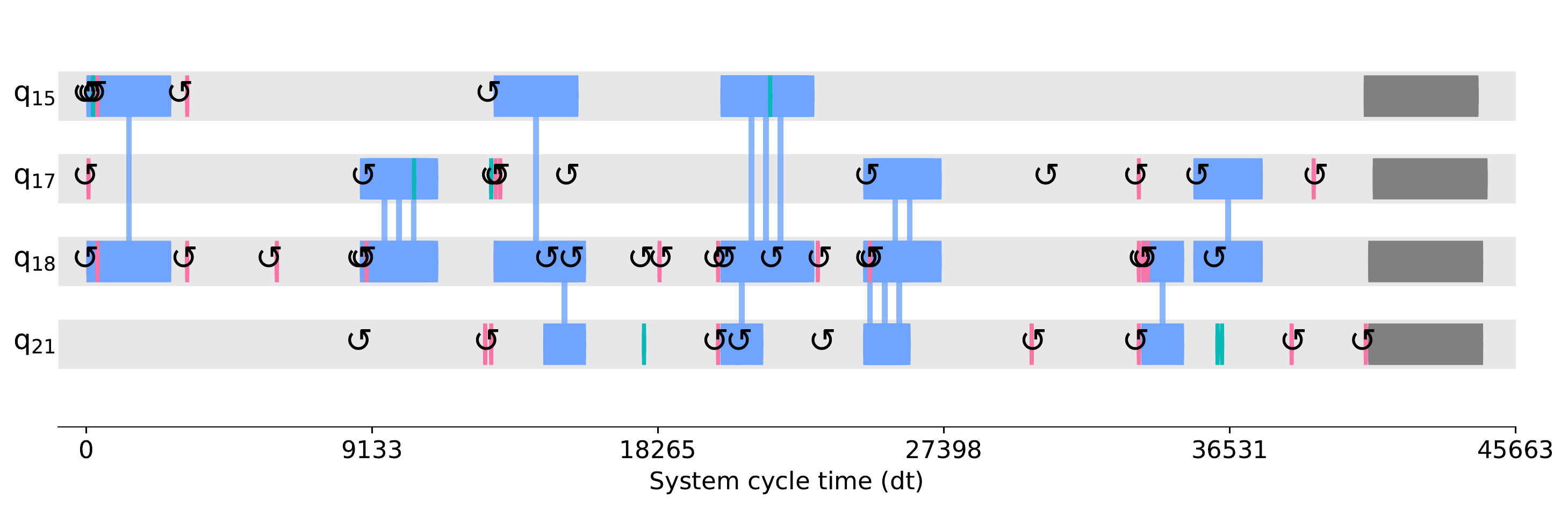}
    \includegraphics[width=0.95\textwidth]{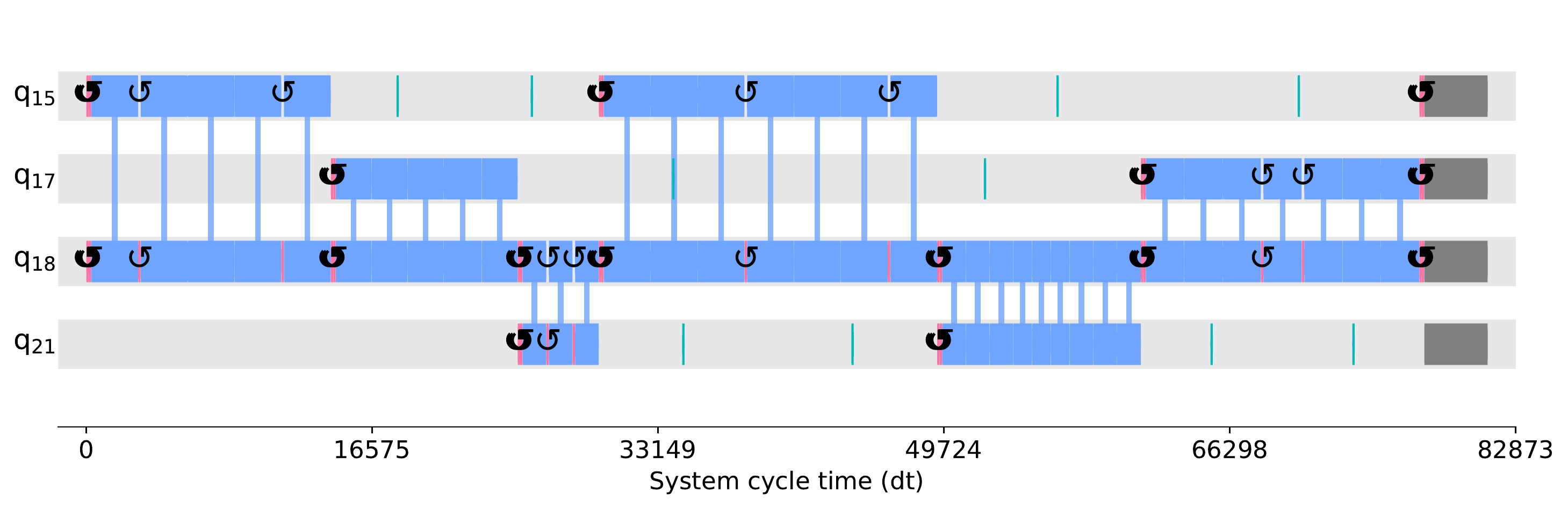}
    \includegraphics[width=0.95\textwidth]{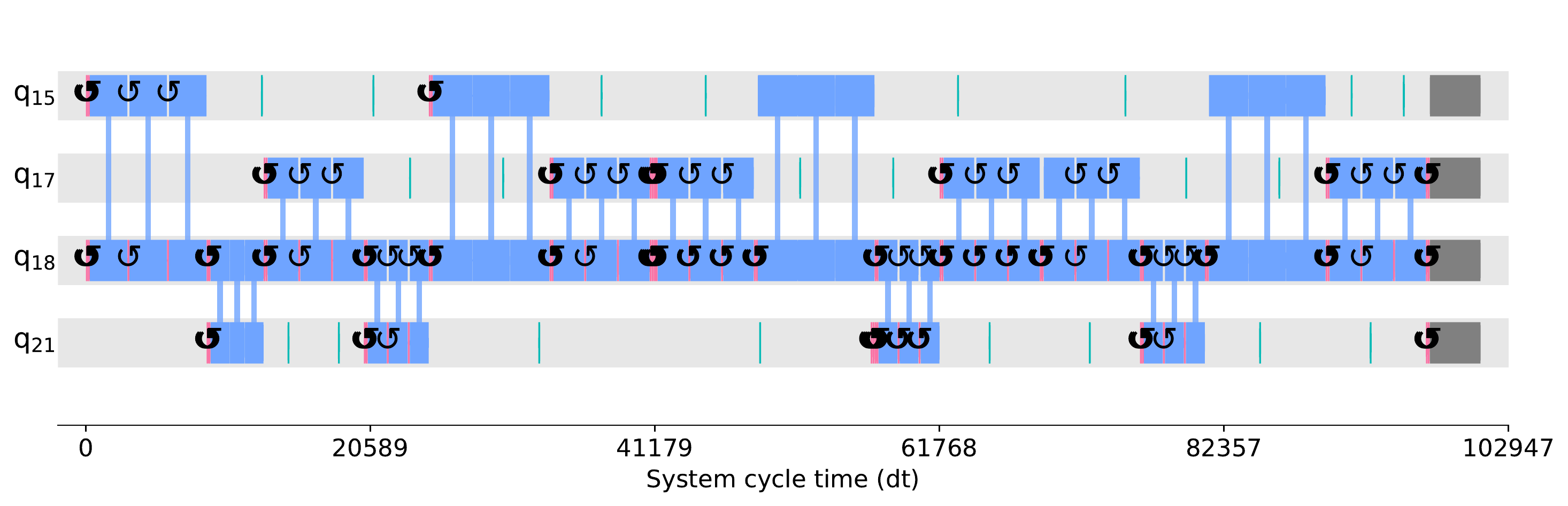}
    \caption{Three Qiskit~\cite{Qiskit} timeline diagrams of a single $n=4$ Quantum Volume circuit compiled to a subgraph of a heavy hex graph, specifically qubits \texttt{15, 18, 17, 21} of \texttt{ibm\_hanoi}. 
    \textbf{Top} diagram shows the compiled QV circuit, with optimizations applied. The \textbf{middle} and \textbf{bottom} circuit diagrams are the same compiled circuit with a digital ZNE circuit folding scale factor of $\lambda=2$ having been applied. Local CNOT circuit folding is used for the \textbf{middle} plot, and global circuit folding is used in the \textbf{bottom} plot. Digital dynamical decoupling sequences of X-X Pauli gates are inserted into all of the the circuits, and the all circuits are scheduled using the ALAP circuit scheduler. 
    The \texttt{RZ} gates are virtual gates~\cite{McKay_2017}, and are represented here by circular black arrow markers. The \texttt{X} gates are shown as green vertical lines, which are very thin because the single qubit gate operations take a small amount of time. Similarly, the short duration single qubit \texttt{SX} gate is represented as vertical red lines. The \texttt{CX} (e.g. CNOT) gates are drawn as vertical blue connections between adjacent qubits in the hardware graph. The dark grey segments at the end of each qubit line denote the measurement of the $4$ qubits. 
    The uncompiled QV circuit contained 24 \texttt{CX} gates, which was then turned into 18 \texttt{CX} gates when optimized using the Qiskit transpiler and adapted to the heavy hex graph structure (shown in the top circuit diagram). ZNE global folding (bottom) with $\lambda=2$ generated a circuit invariant with $36$ \texttt{CX} gates, and local random folding (middle) with $\lambda=2$ generated a circuit with $42$ \texttt{CX} gates. }
    \label{fig:timeline_QV_ZNE_circuits}
\end{figure}

\subsection{Insertion of dynamical decoupling sequences}

Next, dynamical decoupling X-X sequences are inserted into the compiled circuits using the \texttt{PadDynamicalDecoupling}~\cite{PadDynamicalDecoupling} class of the Qiskit~\cite{Qiskit} library. Dynamical Decoupling is an open loop quantum control error suppression technique for mitigating decoherence on idle qubits~\cite{PhysRevLett.82.2417, PhysRevA.58.2733, PhysRevLett.112.050502, 9872062, Kim_2023, https://doi.org/10.48550/arxiv.2207.03670, sym15010062}, which has been shown to help increase measured quantum volumes in previous experiments~\cite{QV_64_IBMQ}. When the dynamical decoupling sequences are inserted into the circuit, the resulting scheduled circuit (based on the IBM Quantum device gate timings) is scheduled according to the ALAP (as late as possible) algorithm.  This dynamical decoupling step is optional, but the intention is to increase the measured heavy output probability of the QV circuits (both with and without noise scaling).

Figure~\ref{fig:timeline_QV_ZNE_circuits} shows two examples of ZNE folded QV circuits represented as circuit timelines, with pairs of Pauli X gates (X-X sequence) dynamical decoupling sequences inserted. Note that the \texttt{rz} gates in Figure~\ref{fig:timeline_QV_ZNE_circuits} are \emph{virtual gates}~\cite{McKay_2017}, meaning that they are performed at the software level and therefore have an error rate of $0$ and a time cost within the circuit of $0$.

\subsection{Execution on hardware and extrapolation to the zero-noise limit}

Once the circuits have been compiled, all of the circuits are submitted to the target backend. Importantly, the number of shots used for the different types of circuits differs---in particular the aim is for the original unscaled (i.e. $\lambda=1$) QV circuits to sample more shots compared to the folded QV circuits. The original QV circuits are all executed with $10000$ shots per circuit. The global folded QV circuits are executed with $1000$ shots per (folded) circuit, and the local randomly folded QV circuits are executed with $100$ shots per (folded) circuit. Finally, the quantum circuit measurements are converted into heavy output proportion (HOP) statistics, and the zero noise extrapolations can be performed on measured HOP data for each individual circuit using Mitiq~\cite{LaRose2022mitiqsoftware}.

In particular, linear~\cite{Temme_2017, LaRose2022mitiqsoftware, Giurgica_Tiron_2020}extrapolation is used, since Richardson extrapolation can lead to extrapolation instability for a limited number of shots. This extrapolation is very computationally efficient. For consistency, and in order to obtain a very robust sampleset, $1000$ random QV circuits are generated and executed for each target backend (and for each target qubit subset of the chip). This number of shots ensures that the resulting extrapolations, no matter what specific scale factors are used, will utilize fewer shots compared to the original (unmodified) QV circuit which is always sampled with $10000$ shots. Therefore, the total overhead that ZNE incurs for this procedure is given by the number of scale factors (not equal to $1$) that are used, which is at most $4$, and each folded-circuit scale factor requires a total of $1,000$ shots to be readout on the device.

With the aim of executing the ZNE QV circuits on the target IBM Quantum devices, the circuits were split into groups of $250$ circuits to be executed together as \emph{jobs} since the maximum number of circuits per job on the target IBM Quantum devices was $300$. This means there were $4$ jobs executed for each group of $1000$ QV circuits, which was then repeated for each of the circuit folding scale factors and the two circuit folding methods. This is important because it is possible for the results to not have been executed together sequentially, and therefore there can be error drift. Additionally, when there are backend device errors and the jobs need to be re-executed, those results would be executed at a different time as well. The results are then post-processed using linear extrapolation, which is very computationally efficient. The ZNE QV results shown in Section~\ref{section:results} are presented as cumulative heavy output probability plots, which are a convenient method of visually representing quantum volume results~\cite{QV_in_practice, QV_64_IBMQ}.

\begin{figure}[th!]
    \centering
    \includegraphics[width=0.8\textwidth]{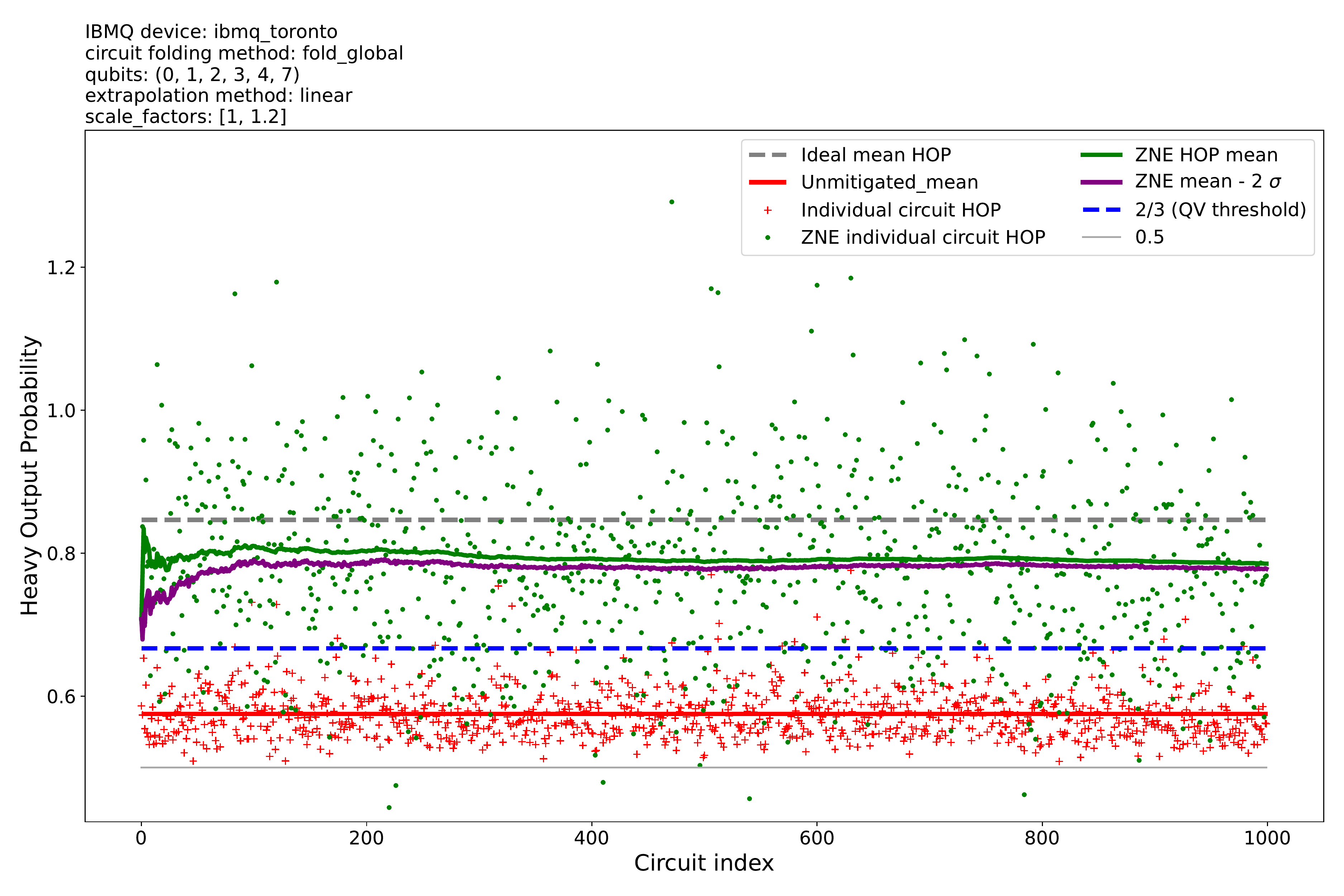}
    \caption{Cumulative HOP plot showing \texttt{ibmq\_toronto} passing the effective quantum volume of $n=6$ for ZNE using unitary circuit folding. The HOP metric is defined to be within $[0, 1]$, note therefore that any ZNE HOP points above $1$ correspond to non-physical extrapolations due to shot noise. }
    \label{fig:main_ibmq_toronto_passing}
\end{figure}

\section{Results}
\label{section:results}

\begin{figure}[h!]
    \centering
    \includegraphics[width=0.8\textwidth]{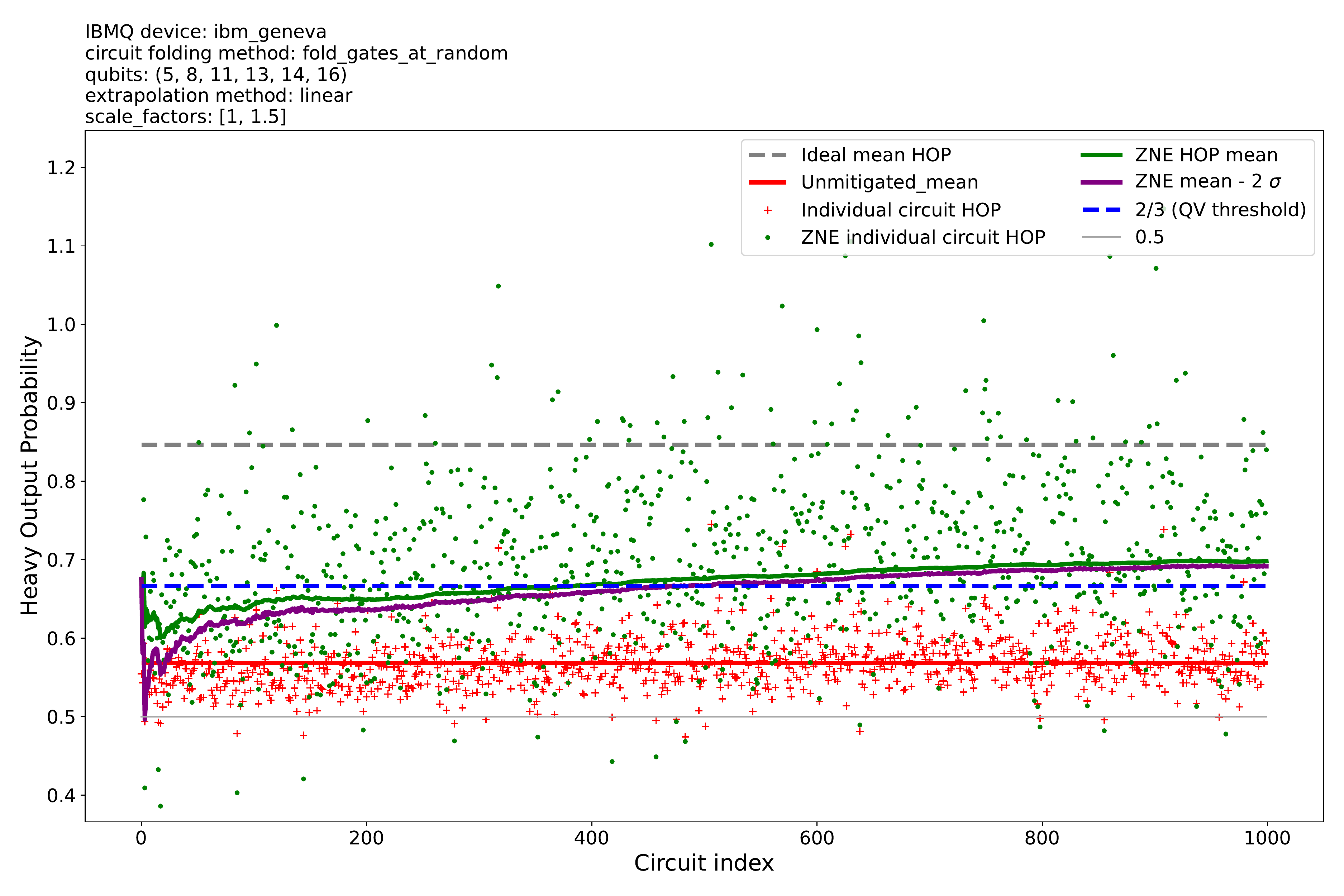}
    \caption{Cumulative HOP plot showing \texttt{ibm\_geneva} passing the effective quantum volume of $n=6$ for ZNE using local random circuit folding. }
    \label{fig:main_ibm_geneva_passing}
\end{figure}

\begin{figure}[h!]
    \centering
    \includegraphics[width=0.8\textwidth]{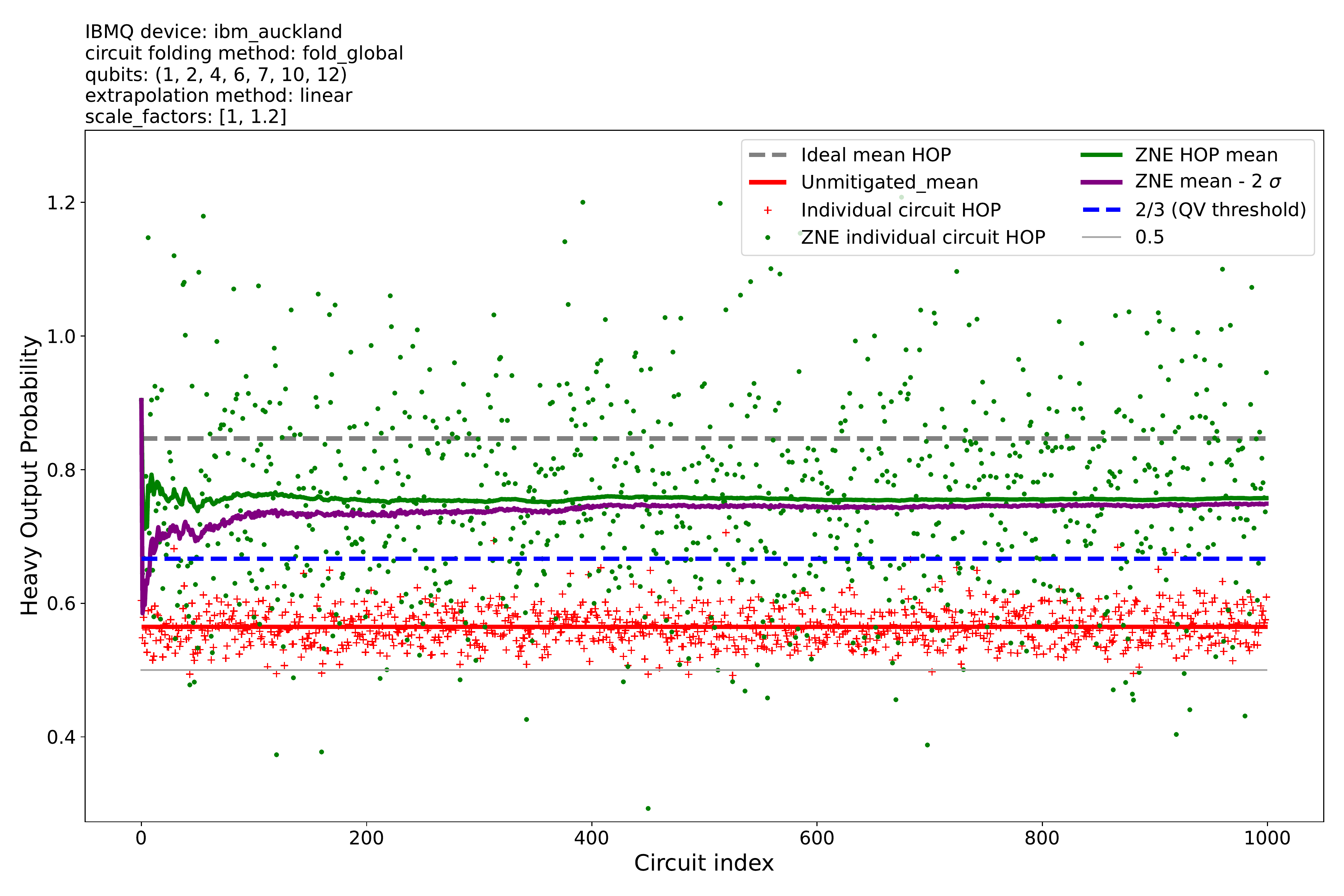}
    \caption{Cumulative HOP plot showing \texttt{ibm\_auckland} passing the effective quantum volume of $n=7$ for ZNE using global circuit folding. }
    \label{fig:main_ibm_auckland_passing}
\end{figure}

\begin{figure}[h!]
    \centering
    \includegraphics[width=0.8\textwidth]{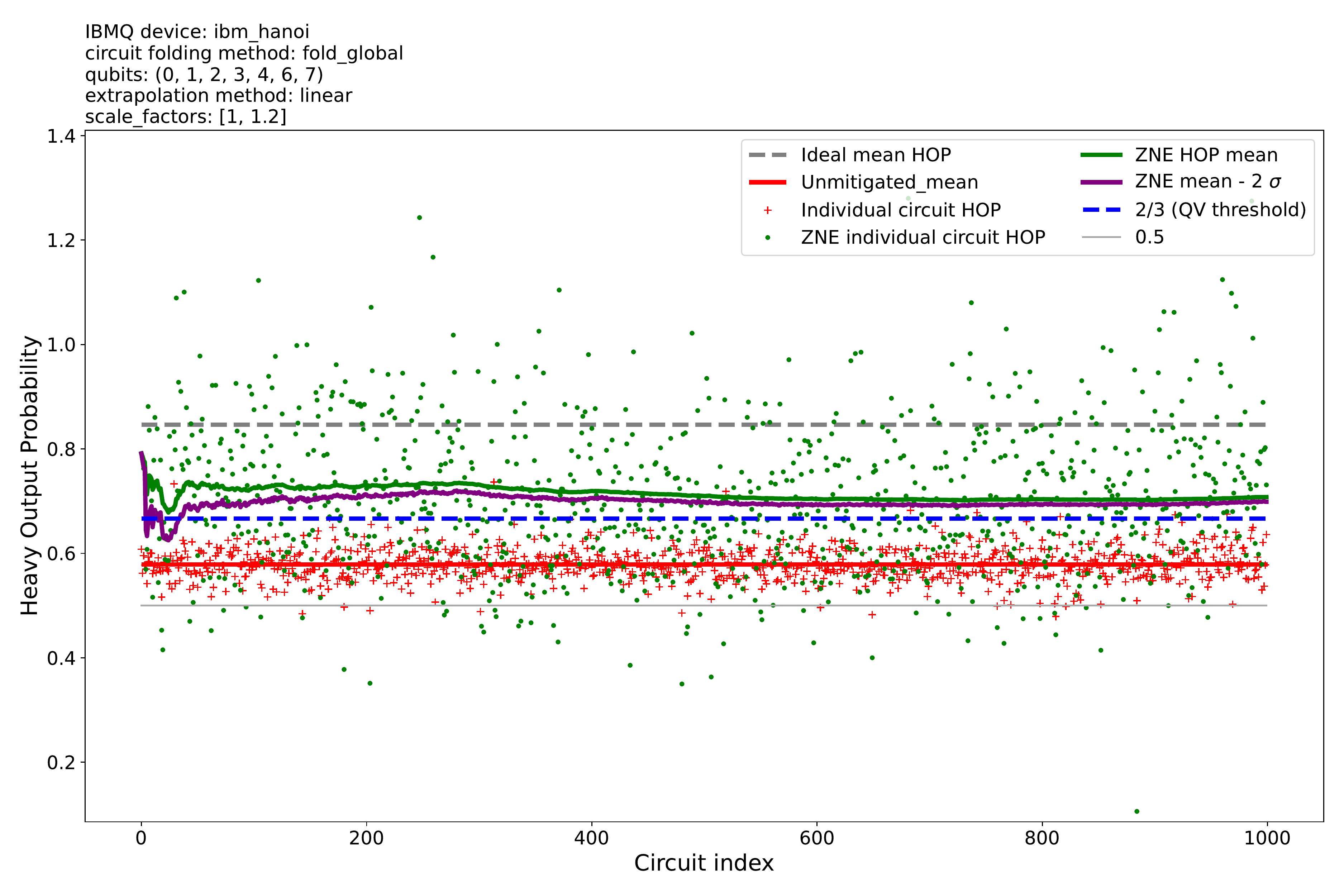}
    \caption{Cumulative HOP plot showing \texttt{ibm\_hanoi} passing the effective quantum volume of $n=7$ for ZNE using unitary circuit folding. }
    \label{fig:main_ibm_hanoi_passing}
\end{figure}

The effective quantum volume results are reported in the form of cumulative heavy output probability plots. The $2 \sigma$ bound on the error mitigated HOP values are measured using bootstrapped re-sampling~\cite{Baldwin2022reexaminingquantum, effective_QV} on the (cumulative) ZNE HOP points. The re-sampling is performed $100$ times for the full HOP vector size (of $1000$), with replacement, and then the standard deviation $\sigma$ is computed. Figures~\ref{fig:main_ibmq_toronto_passing} and~\ref{fig:main_ibm_geneva_passing} show \texttt{ibmq\_toronto} and \texttt{ibm\_geneva} both passing an effective quantum volume of $n=6$, respectively. Figures~\ref{fig:main_ibm_auckland_passing} and~\ref{fig:main_ibm_hanoi_passing} show \texttt{ibm\_auckland} and \texttt{ibm\_hanoi} passing an effective quantum volume of $n=7$, respectively. Each of the four figures, show an  effective quantum volume test averaged over $1000$ circuits passing the $\frac{2}{3}$ threshold. Therefore, each plot show that the effective Quantum Volume is increased by ZNE beyond the vendor-measured quantum volume, a fact that was never been demonstrated to date on a NISQ computer~\cite{effective_QV}. It is important to consider however, that although effective quantum volume is defined, as was the original Quantum Volume metric~\cite{Quantum_Volume}, as the \emph{largest} such measurement that can be made on a device, it is the case that a wide range of QV results can be observed on a fixed NISQ computer~\cite{QV_in_practice} depending on the choice of qubits and error mitigation method. In the interest of examining in what cases we observed effective quantum volume experiments failing to pass the QV threshold, we report the associated negative results in Appendices~\ref{section:appendix_ZNE_unable_to_extrapolate},~\ref{section:appendix_ZNE_QV_not_passing_threshold}. Moreover, Appendix~\ref{section:appendix_compare_unitary_and_local_circuit_folding} briefly gives some additional cumulative ZNE HOP plots to illustrate how similar the two ZNE circuit folding methods are. For all of the experiments we present, linear extrapolation is used because it gives the lowest extrapolation variability (specifically, extrapolated values that are more typically consistent with the physical HOP range). The better stability of linear extrapolation compared to high-order polynomial fittings (e.g. Richardson extrapolation) is a known phenomenon that can be explained in terms of under-fitting vs over-fitting of the noise scaled expectation values (see Ref.~\cite{Giurgica_Tiron_2020} for more details on the extrapolation error).

Figure~\ref{fig:main_ibm_auckland_passing} shows a single cumulative HOP plot for a case where \texttt{ibm\_auckland} passes the effective quantum protocol of size $n=7$ when executing on qubits \texttt{1, 2, 4, 6, 7, 10, 12}. The ZNE protocol in this case used scale factors of $1$ and $1.2$, global circuit folding method, and linear extrapolation on the noise-scaled ZNE data points. Figure~\ref{fig:main_ibm_auckland_passing} shows clearly that the original QV circuits were sampled below $\frac{2}{3}$, but above $0.5$, and that the error mitigated HOP values were measurably above the $\frac{2}{3}$ threshold but not above the ideal mean HOP. Figure~\ref{fig:main_ibm_auckland_passing} also shows that some of ZNE HOP points were measured to be above $1$, which is possibly due to the relatively small shot count, and the fractional scale factors that are used resulting in some extrapolation instability. Extrapolation instability, in particular, extrapolated HOP which is much greater than $1$, is seen in Figures~\ref{fig:main_ibmq_toronto_passing},~\ref{fig:main_ibm_geneva_passing},~\ref{fig:main_ibm_auckland_passing}, and~\ref{fig:main_ibm_hanoi_passing}.

\begin{figure}[h!]
    \centering
    \includegraphics[width=0.49\textwidth]{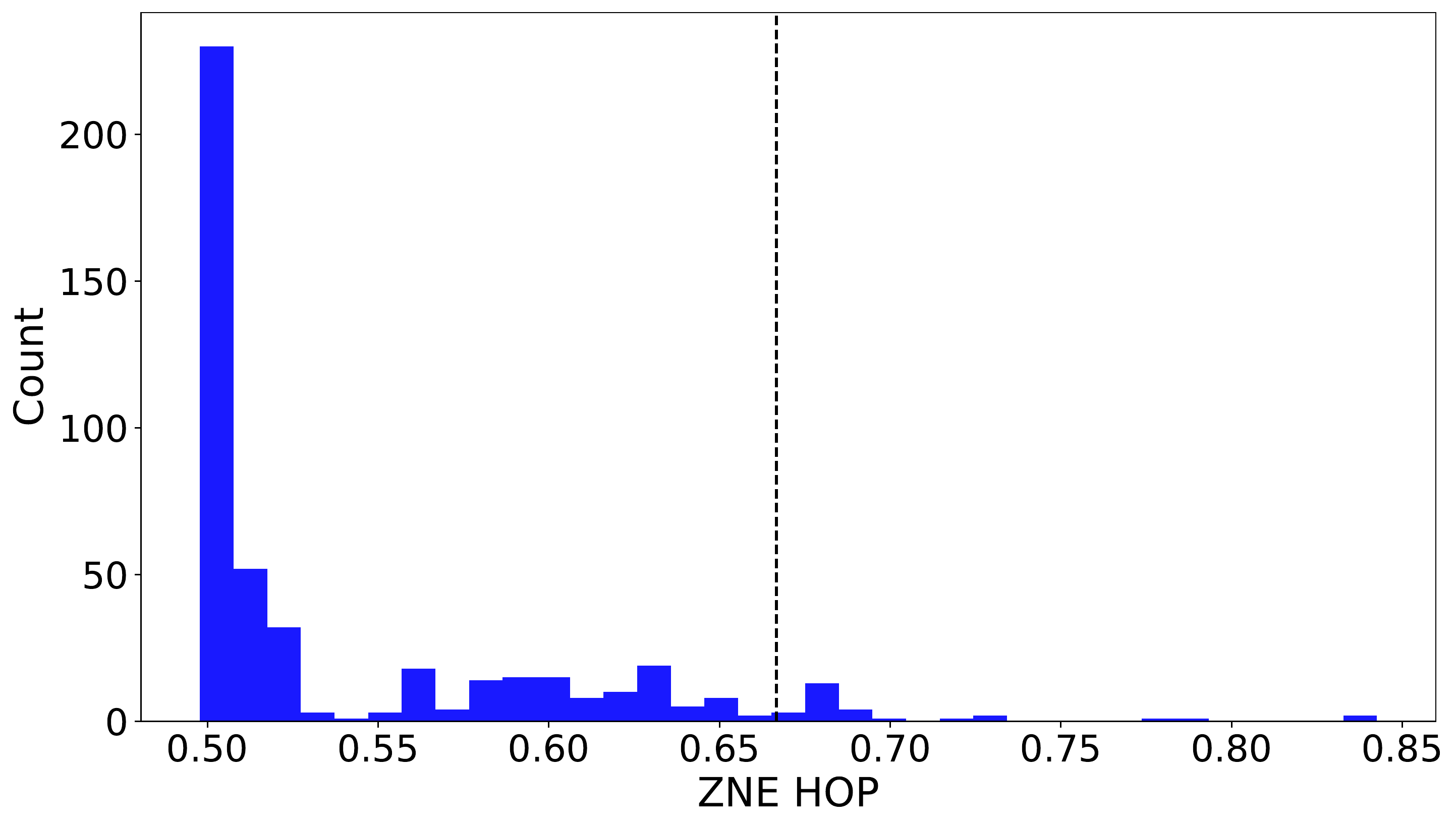}
    \includegraphics[width=0.49\textwidth]{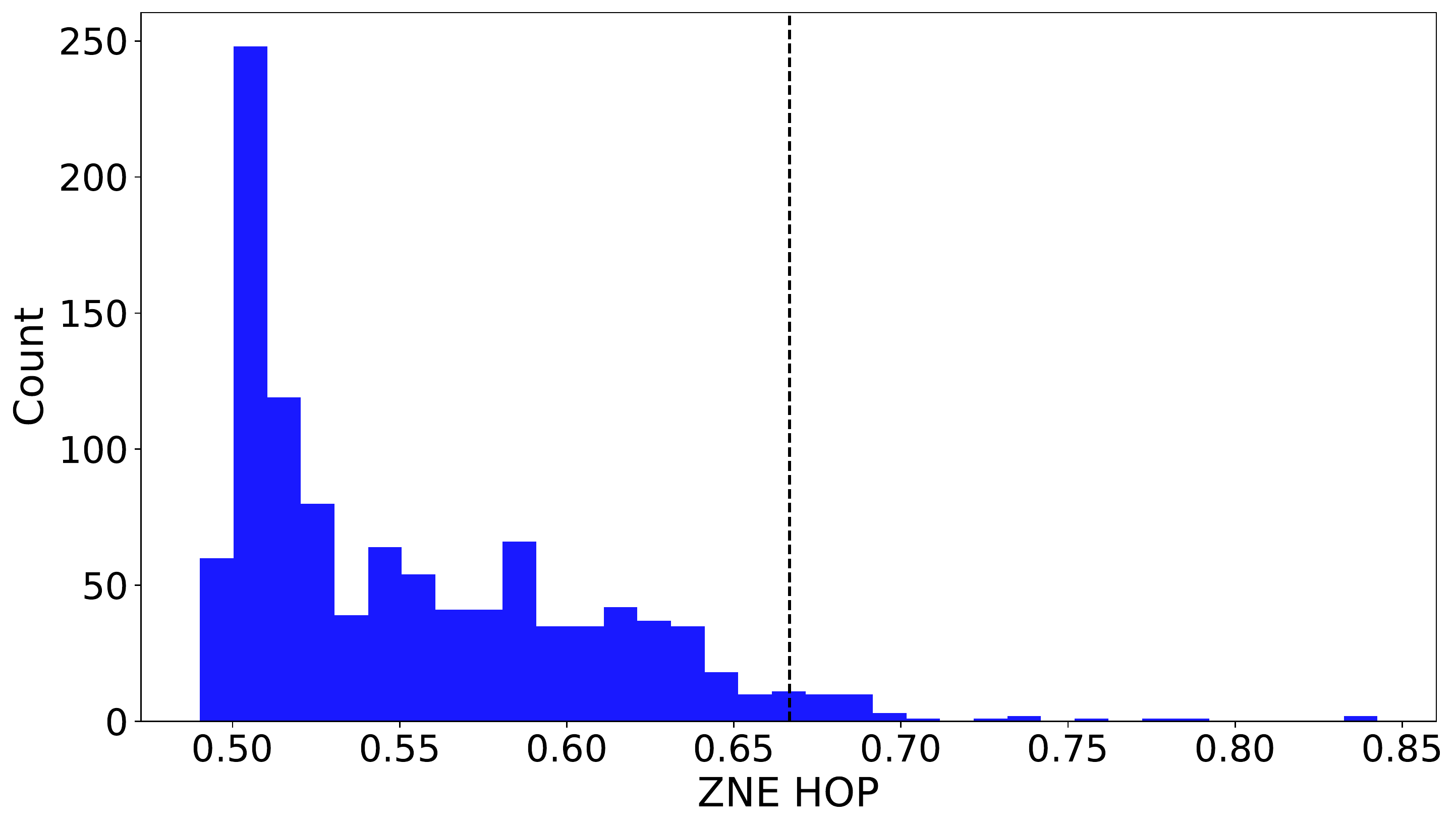}
    \caption{Distribution of all mean $\lambda=0$ Zero Noise Extrapolated HOP values, across all of the different noise scale extrapolations, for $n=6$ (left), run on \texttt{ibmq\_toronto} and \texttt{ibm\_geneva} and $n=7$ (right) run on \texttt{ibm\_hanoi} and \texttt{ibm\_auckland}. The $\frac{2}{3}$ threshold is marked with the vertical dashed black line. }
    \label{fig:ZNE_HOP_histograms}
\end{figure}

Figure~\ref{fig:ZNE_HOP_histograms} shows the full distribution of measured ZNE HOP values (specifically the mean ZNE HOP values across each set of $1000$ QV circuits) across all of the experiments that were executed. This distribution of ZNE HOP values includes the different linear extrapolations for $\lambda=0$ across the possible combinations of the scale factors that were used between $\lambda=1$ and $\lambda=1.2, 1.5, 1.8, 2$. These distributions are not comprehensive for the entire hardware graph of the four IBMQ chips, but instead represent only a random subset of the possible hardware graphs that could be chosen to compile to. As has been observed for Quantum Volume test enumeration~\cite{QV_in_practice}, Figure~\ref{fig:ZNE_HOP_histograms} shows that on average the performance is worse than the best performance cases. Lastly, note that the best mean ZNE HOP does not reach above the ideal mean HOP of $\approx 0.85$~\cite{10.5555/3135595.3135617, Quantum_Volume}, showing that when this procedure is averaged over many instances the quantum error mitigation results in measures that are consistent with the physical measure of HOP. 

An important aspect of zero noise extrapolation is that statistical noise is increased by the extrapolation process. For example, if there are too few measurement shots used for computing an expectation value, or if there is an error rate drift on the quantum computer during the period of time the circuits are executed, the resulting extrapolations may end up outside the expected physical range. For a noiseless quantum computer, individual circuits may have a heavy output probability that is slightly larger or smaller than $\frac{1 + \ln(2)}{2} \approx 0.85$~\cite{Baldwin2022reexaminingquantum, QV_in_practice} (especially if $n$ is small). In our experiments, we observe that ZNE sometimes amplifies such deviations beyond the theoretical upper bound and, for a few cases, the extrapolated HOP is even larger than $1$.
The amplification of statistical fluctuations is a known aspect of ZNE and is expected for individual circuits due to the very small number of sample (100 shots). However, at least when using linear fit (more stable with respect to Richardson extrapolation), when averaging over the 1000 random circuit of the QV protocol, we always obtained a mean HOP within the physical range.

Another important note about ZNE, and specifically ZNE QV, is that if there is no signal for small-scale factors, and especially, $\lambda=1$ then the extrapolation simply does not work. In the case of Quantum Volume, if the measured heavy output probabilities across the scale factors are all around $0.5$ (meaning the computation has completely decohered) then the extrapolation cannot work. Furthermore, even if some of the larger scale factors converge to a HOP of $0.5$ then extrapolations based on those results may also end up skewed---this in part motivated the use of scale factors that were $\leq 2$. For this reason, our experiments target several different parts of the chips since the error rates on each chip can vary significantly~\cite{QV_in_practice}, so as to quantify a \emph{distribution} of the possible ZNE QV performance on the target QPUs. This concept of applying the same algorithm to an ensemble of different parts of the quantum computer hardware has been used in other contexts~\cite{10.1145/3352460.3358257}.

\section{Discussion and Conclusion}
\label{section:conclusion}
We have demonstrated an increase in the measured effective quantum volume over the vendor-benchmarked quantum volume, on four IBM Quantum devices \texttt{ibm\_geneva}, \texttt{ibm\_auckland}, \texttt{ibm\_hanoi}, \texttt{ibmq\_toronto}. These results show that error mitigation can be scaled to larger QV circuit sizes than have been previously measured in the effective quantum volume protocol~\cite{effective_QV}. We have also shown examples where quantum error mitigation, in particular ZNE, is unable to extrapolate the computation to the zero noise limit---this occurs when the circuit that is executed experiences too much decoherence and there is no signal that can be extrapolated. Because the QV ZNE protocol fails to perform a meaningful zero noise extrapolation when all of the computations have decohered because of errors in the computation, effective quantum volume serves as a meaningful quantification of how effective quantum error mitigation can be for near-term quantum processors. In particular, although the effective quantum volume can be greater than the vendor-measured quantum volume (as has been demonstrated in this study), there is a hard limit where the extrapolation will not work (when the mean heavy output probability converges to $0.5$) which is governed by the error rate of the quantum computer. Therefore, as with the original quantum volume test, effective quantum volume is a difficult test for near-term quantum processors to pass, and importantly as a benchmark employs the full stack of software, hardware, and algorithms involved with utilizing quantum error mitigation.

Probing how other quantum error mitigation techniques, in particular, probabilistic error cancellation (PEC)~\cite{Zhang_2020, PhysRevApplied.15.034026, Temme_2017}, perform when applied to effective Quantum Volume is an open question. In general, other quantum error suppression mechanisms such as Pauli twirling (also known as randomized compiling)~\cite{PhysRevA.94.052325, ferracin2019accrediting, cai2019constructing, jain2023improved} and variants of dynamical decoupling should continue to be analyzed for their effectiveness at improving NISQ computation quality, and therefore also how they impact full stack benchmarks such as Quantum Volume.

\section{Acknowledgments}
\label{section:acknowledgments}
This work was supported by the U.S. Department of Energy through the Los Alamos National Laboratory. Los Alamos National Laboratory is operated by Triad National Security, LLC, for the National Nuclear Security Administration of U.S. Department of Energy (Contract No. 89233218CNA000001). The research presented in this article was supported by the Laboratory Directed Research and Development program of Los Alamos National Laboratory under project number 20220656ER, and by the NNSA's Advanced Simulation and Computing Beyond Moore's Law Program at Los Alamos National Laboratory. This research used resources provided by the Los Alamos National Laboratory Institutional Computing Program, which is supported by the U.S. Department of Energy National Nuclear Security Administration under Contract No. 89233218CNA000001. This work has been assigned LANL report number LA-UR-23-26260. We acknowledge the use of IBM Quantum services for this work. The views expressed are those of the authors and do not reflect the official policy or position of IBM or the IBM Quantum team. 

\noindent
This work was supported by the U.S. Department of Energy, Office of Science, Office of Advanced Scientific Computing Research, Accelerated Research in Quantum Computing under Award Numbers DE-SC0020266 and DESC0020316 as well as by IBM under Sponsored Research Agreement No. W1975810. AM acknowledges support from the PNRR MUR project PE0000023-NQSTI (Italy). 

\noindent
The figures in this study are generated using a combination of matplotlib~\cite{thomas_a_caswell_2021_5194481, Hunter:2007}, networkx~\cite{hagberg2008exploring}, Quantikz~\cite{quantikz}, and Qiskit~\cite{Qiskit} in Python 3.

\appendix

\section{ZNE unable to extrapolate if the computation contains too many errors}
\label{section:appendix_ZNE_unable_to_extrapolate}

\begin{figure}[h!]
    \centering
    \includegraphics[width=0.49\textwidth]{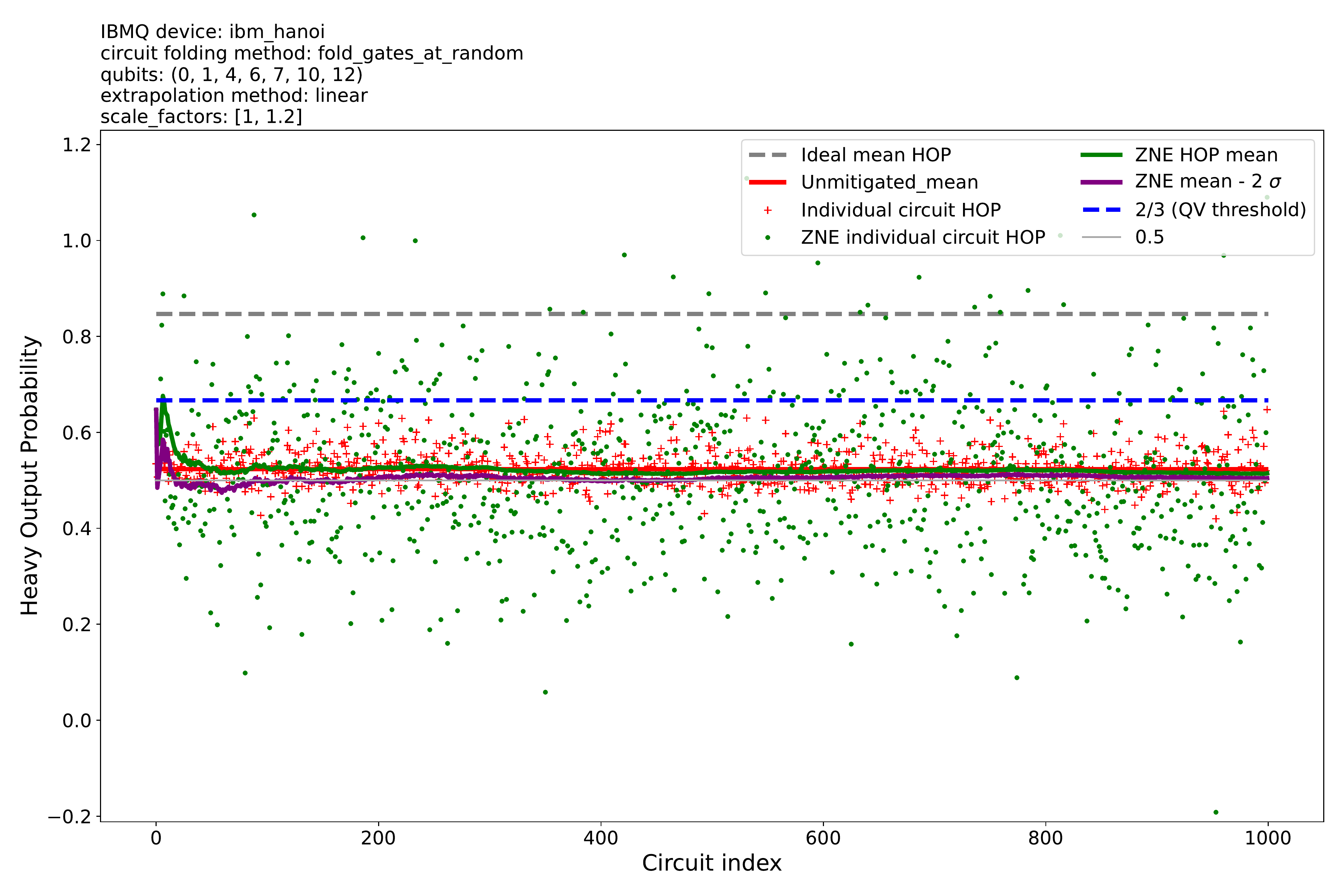}
    \includegraphics[width=0.49\textwidth]{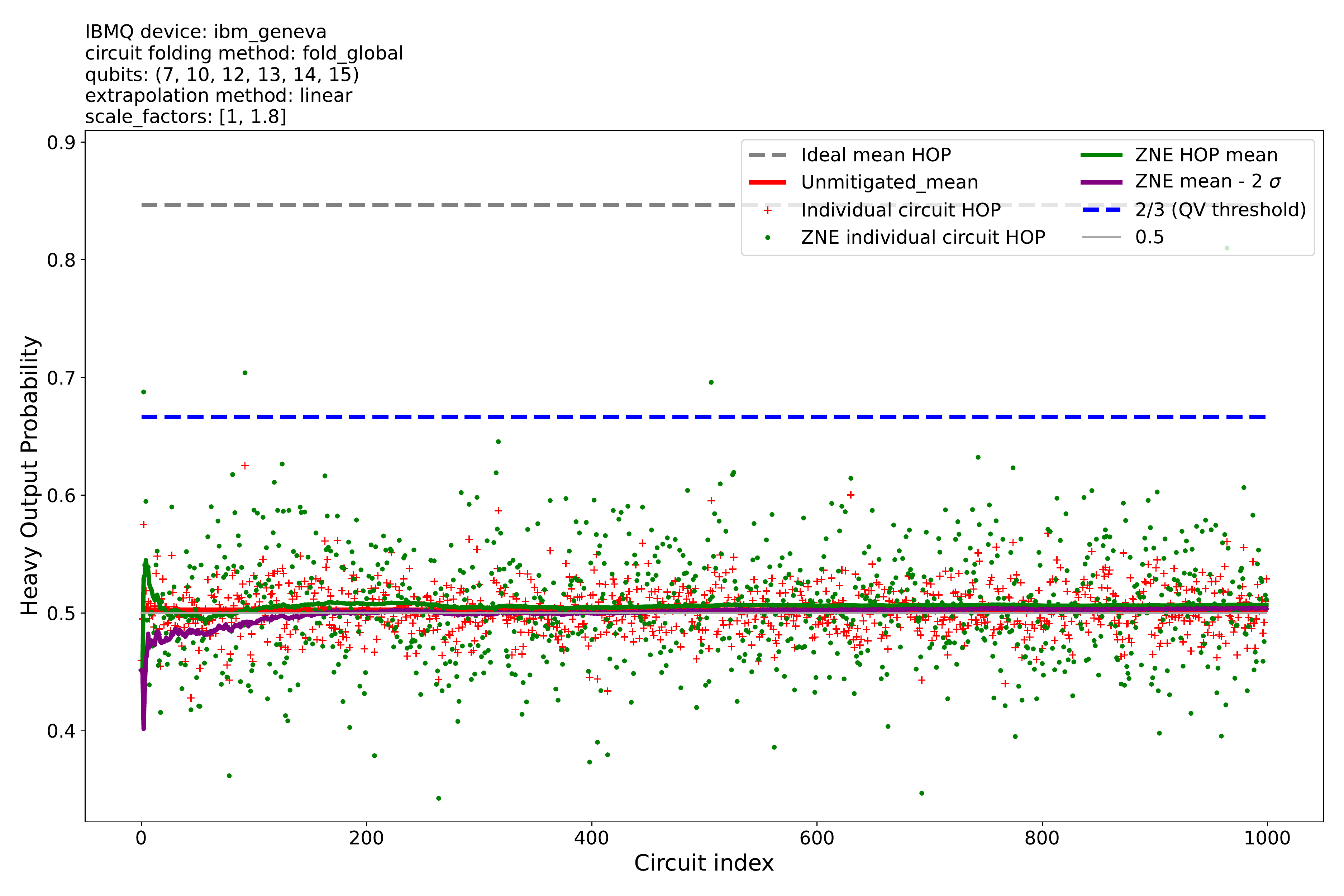}
    \includegraphics[width=0.49\textwidth]{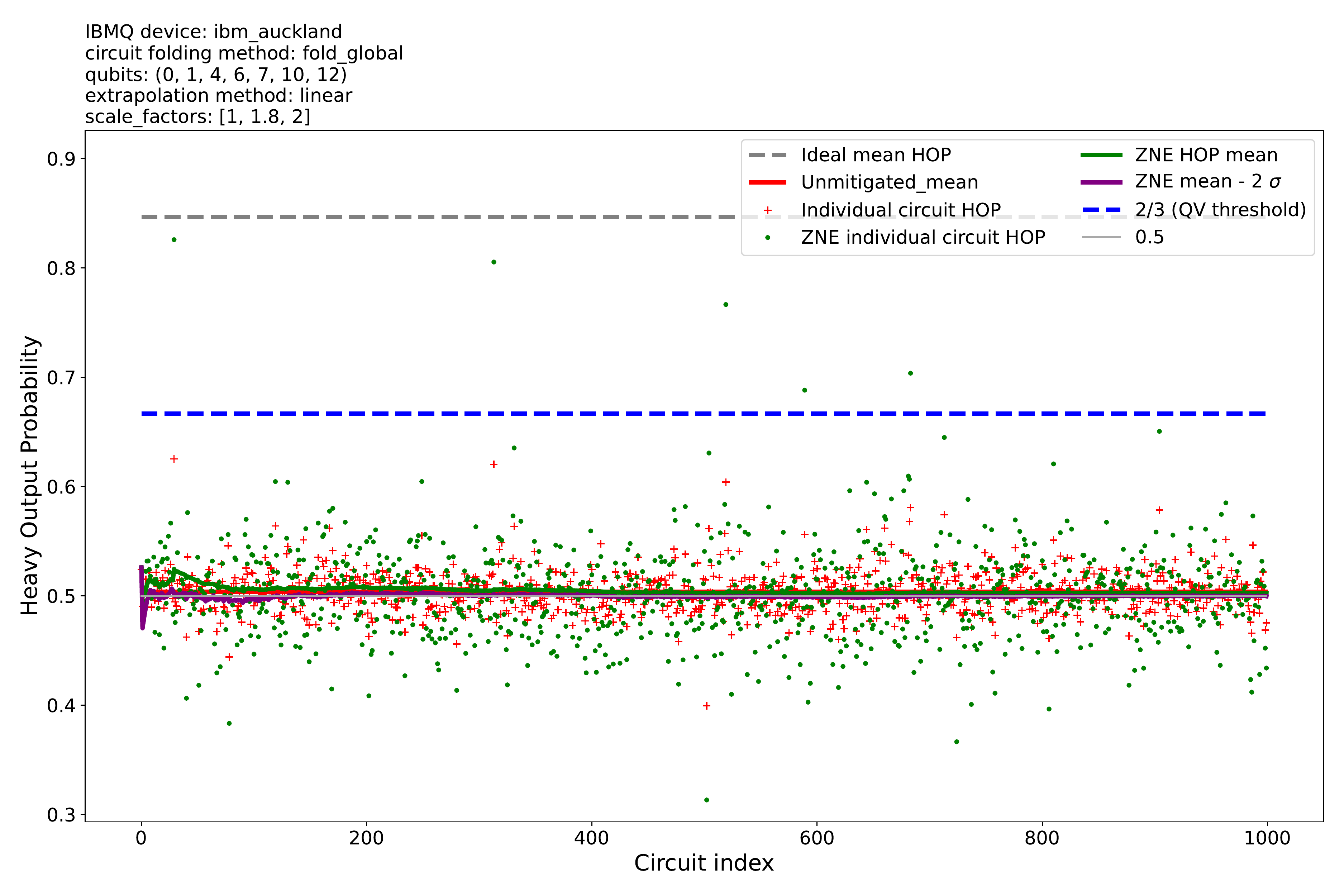}
    \includegraphics[width=0.49\textwidth]{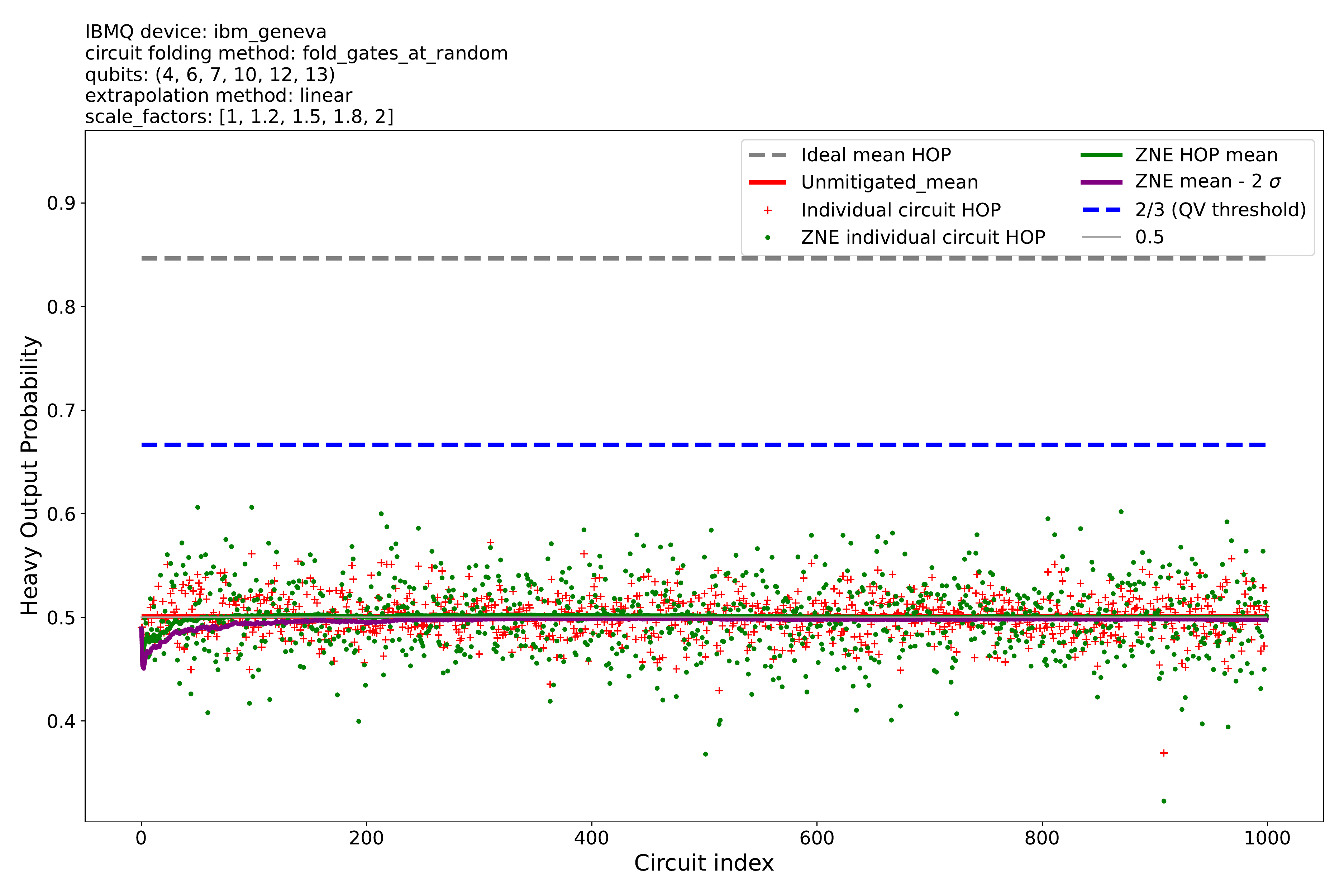}
    \caption{Four example cumulative HOP plots which show instances where the total errors encountered within the circuits caused the mean HOP to converge to $0.5$, which is the HOP that is expected when all fidelity in the computation is lost. This high noise convergence to $0.5$ HOP has been observed before~\cite{QV_in_practice}, but here it is notable because it means that any extrapolation of expectation values would not work, since the extrapolations would be for a flat expectation value landscape (of $0.5$ on average). Experiment parameters are shown in the plot titles. }
    \label{fig:appendix_HOP_ZNE_converged_0.5}
\end{figure}

Figure~\ref{fig:appendix_HOP_ZNE_converged_0.5} shows examples from the set of experiments we executed where the original unmodified QV circuits (e.g. $\lambda=1$) were sampled with a mean HOP of $0.5$. The result of this is that any extrapolation for increased error can not work. This occurs in general when the computation reaches the maximal amount of noise, or error. This shows the limits of ZNE, for the specific case of Quantum Volume. In particular, this shows that such near term quantum error mitigation algorithms require at least some amount of signal to be present in order for the error mitigation to work.

\section{Compare global and local circuit folding}
\label{section:appendix_compare_unitary_and_local_circuit_folding}

\begin{figure}[h!]
    \centering
    \includegraphics[width=0.49\textwidth]{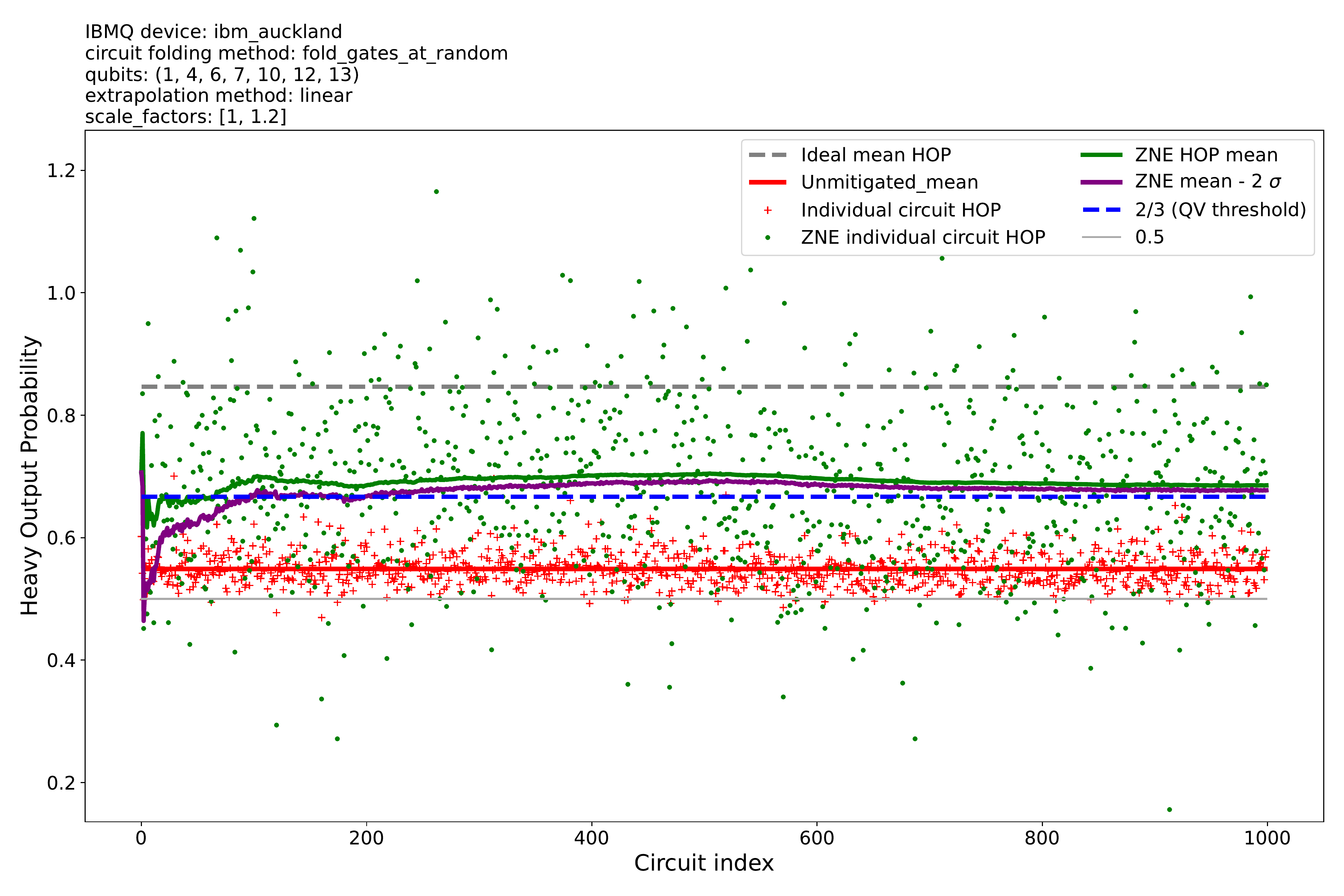}
    \includegraphics[width=0.49\textwidth]{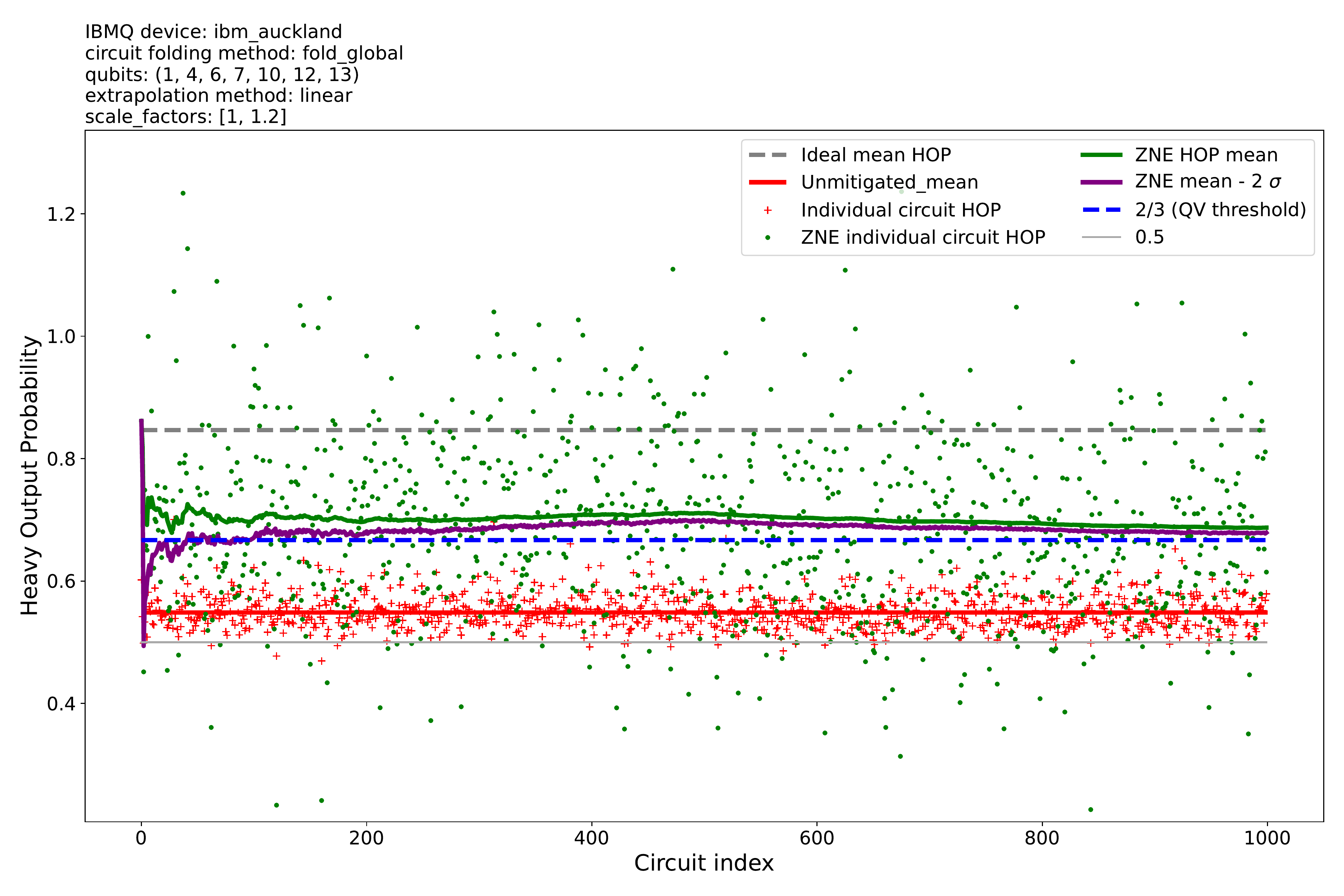}
    \includegraphics[width=0.49\textwidth]{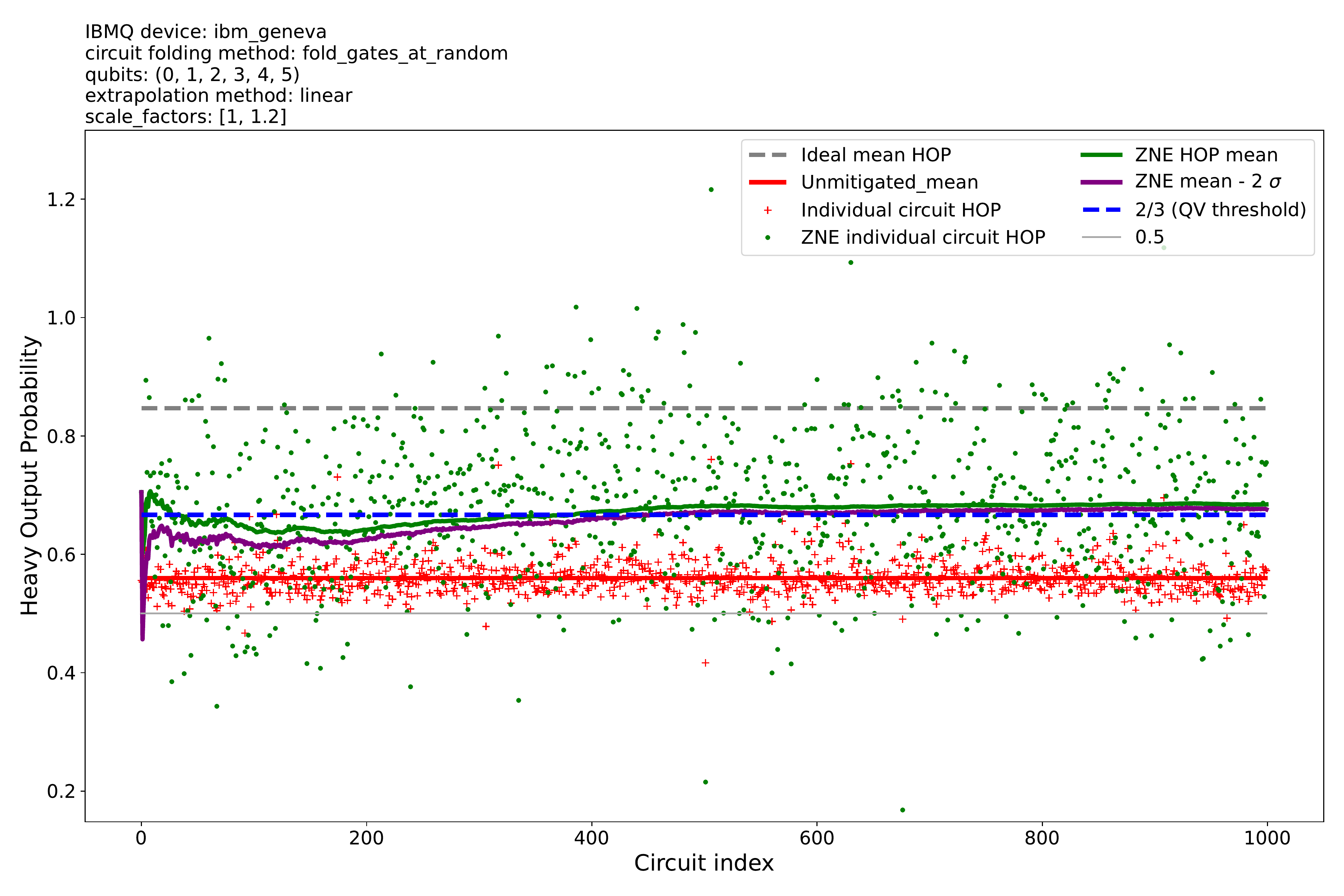}
    \includegraphics[width=0.49\textwidth]{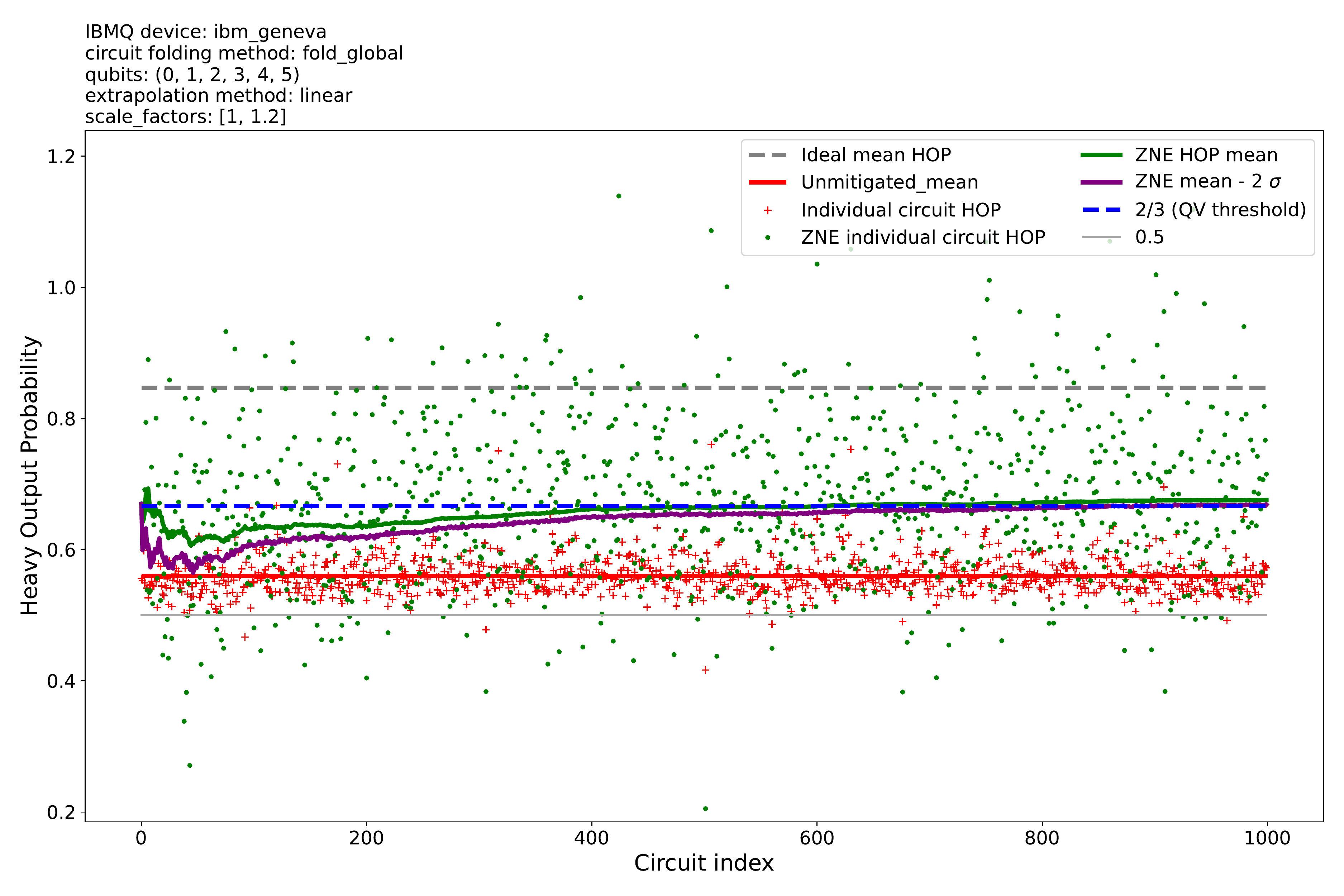}
    \caption{Comparison of global folding (right column) and local random circuit folding (left column) HOP results with all other parameters being held constant. Experiment parameters are shown in the plot titles. }
    \label{fig:appendix_unitary_local_circuit_folding}
\end{figure}

Both global and local folding produce the same intended effect, which is to increase the effective noise of a circuit by intentionally increasing the number of gates. However, the circuit implementations of these two folding methods are not identical, and can be compiled to devices in different ways. In particular the way the fractional scale factors are implemented is different for these two methods. Generally, sampling more random instances of locally folded circuits with fractional scale factors is useful because it can average out biases from increased noise in local parts of the circuit. With global folding instead the process is deterministic but only folds part of the circuit unitary, which means the results could have biased noise. Given these differences, a relevant question is whether the two methods performed similarly in the effective quantum volume experiments considered in this work. Overall, the two methods did perform comparably. Figure~\ref{fig:appendix_unitary_local_circuit_folding} shows some side-by-side comparisons of the measured HOP extrapolation plots which show that both methods did perform similarly. Specifically, these plots show two instances of identical parameter settings (same IBM Quantum device, qubits, layout, scale factors, and extrapolation algorithm), where only the choice of circuit folding method was changed. Although it is not possible to perfectly control the time within which these circuits were executed, these circuits were queued in sequence with each other, which means that time was controlled for as well as much as it can be on an IBM Quantum system.

\section{ZNE QV not passing threshold}
\label{section:appendix_ZNE_QV_not_passing_threshold}

\begin{figure}[h!]
    \centering
    \includegraphics[width=0.49\textwidth]{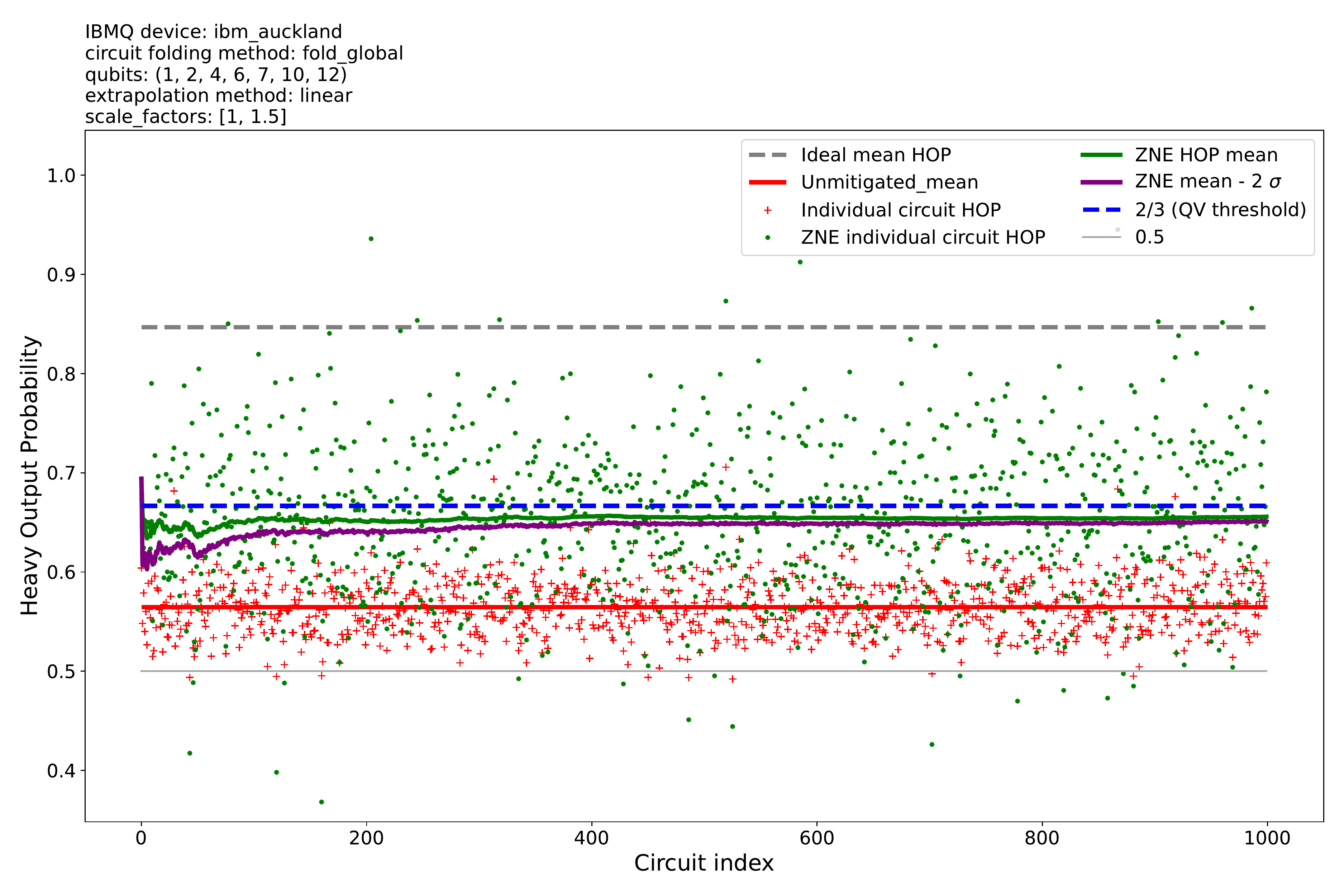}
    \includegraphics[width=0.49\textwidth]{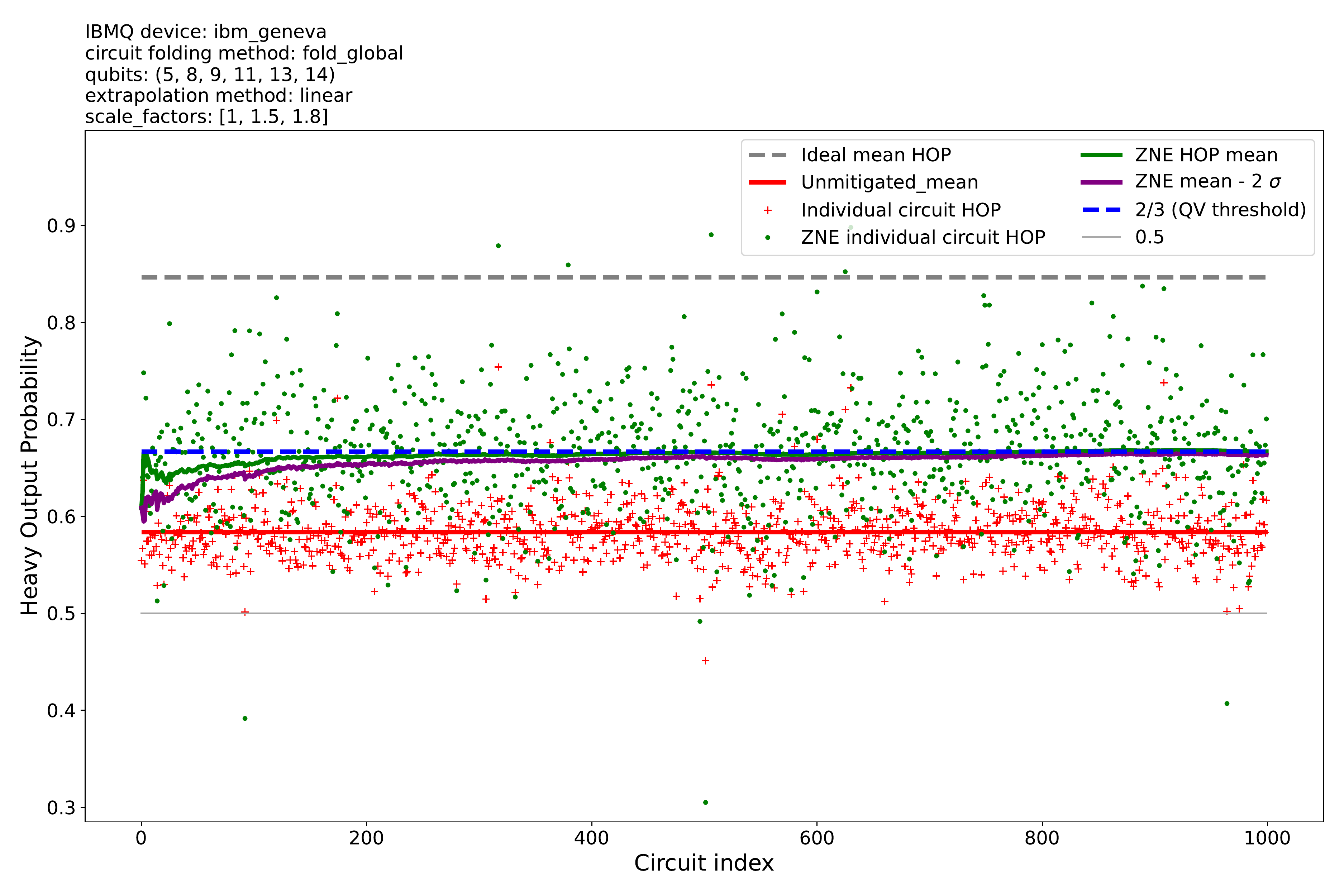}
    \includegraphics[width=0.49\textwidth]{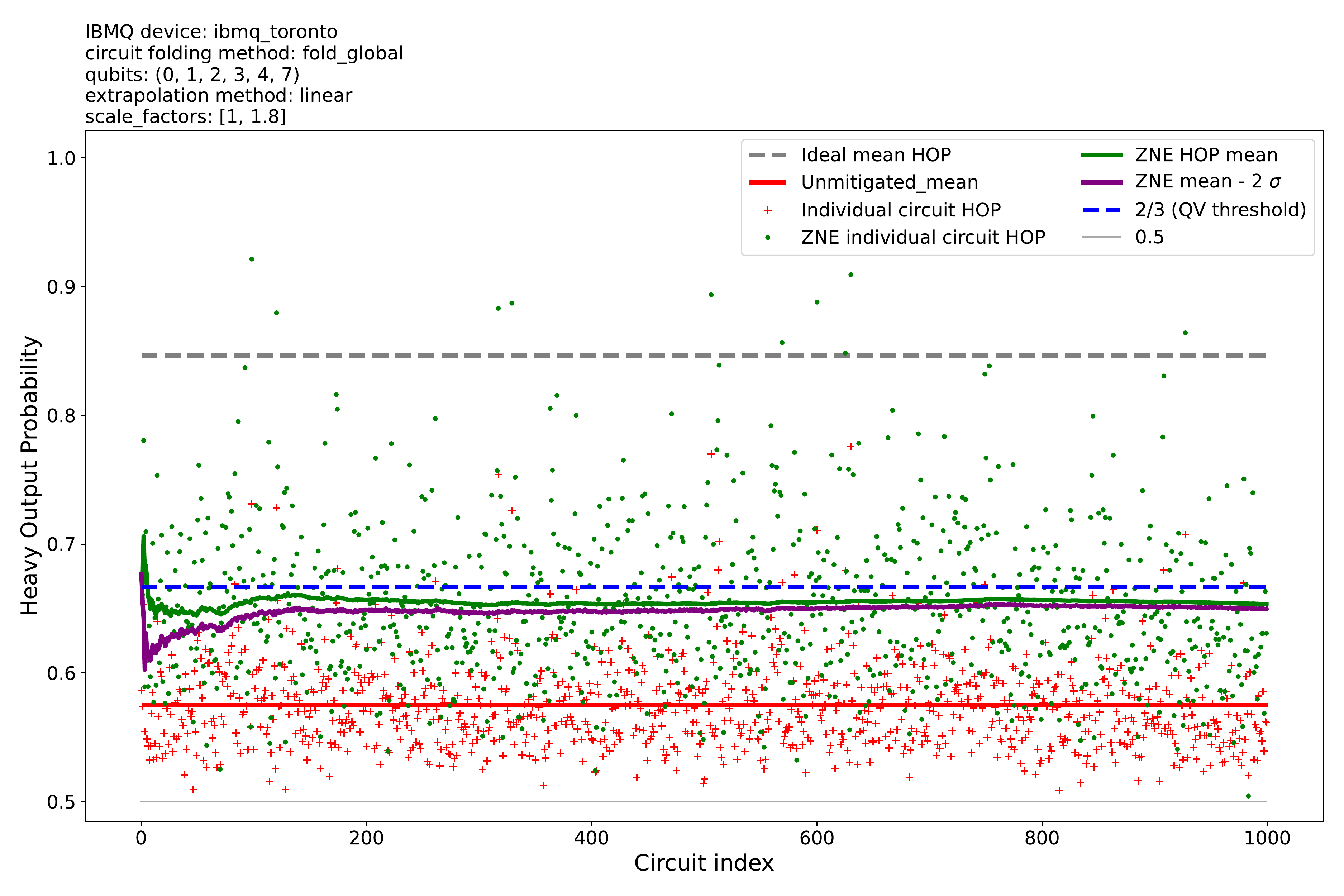}
    \includegraphics[width=0.49\textwidth]{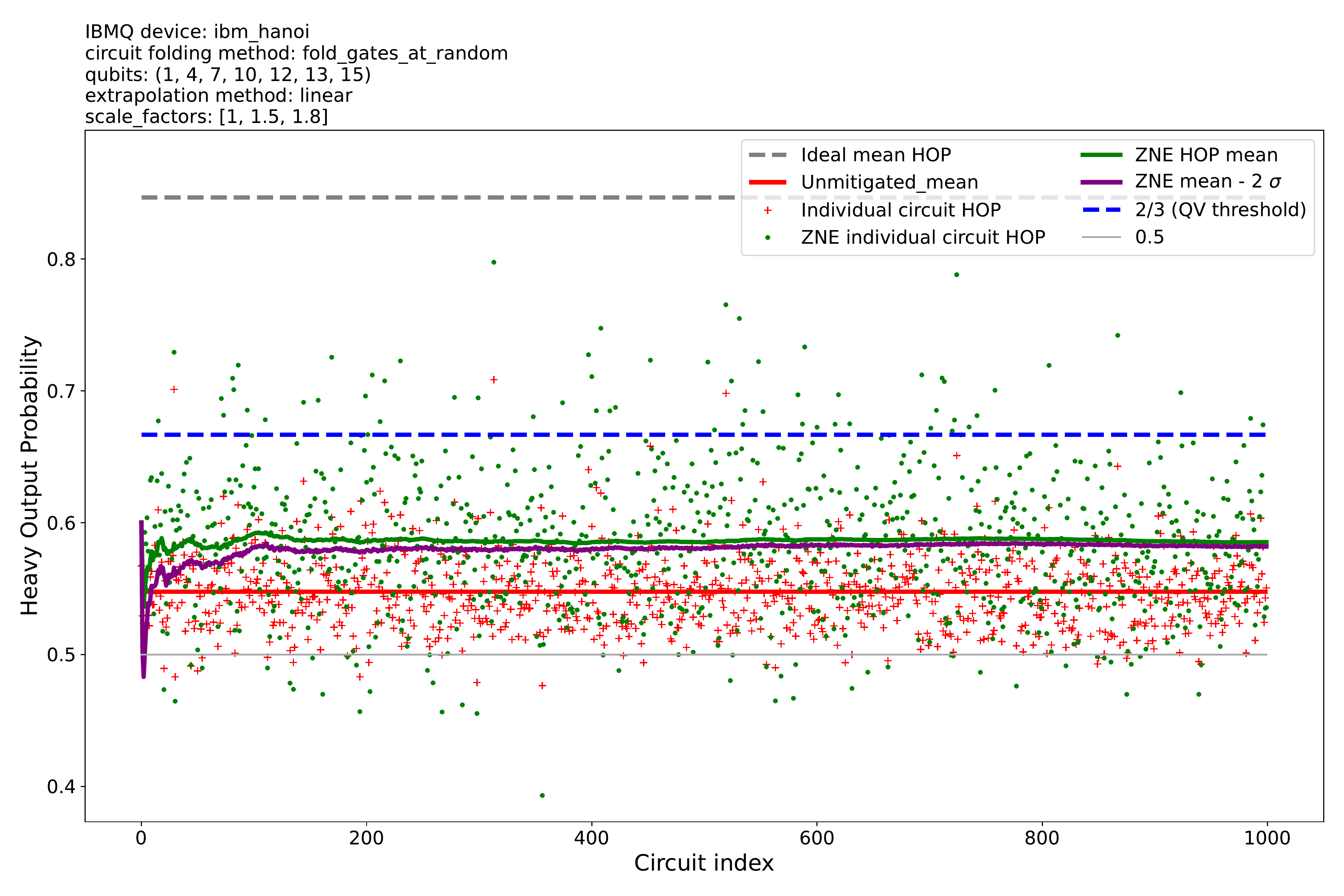}
    \caption{Example plots of the four IBM Quantum devices failing to pass the $\frac{2}{3}$ threshold with high confidence for the ZNE QV protocol under various ZNE settings and IBM Quantum compilation qubit subgraphs. Experiment parameters are shown in the plot titles. }
    \label{fig:example_QV_not_passing_threshold}
\end{figure}

Like what was observed on numerous IBM Quantum devices in ref.~\cite{QV_in_practice}, for this set of ZNE QV tests evaluated in this study, many of the settings (in particular, the parts of the chip that were executed on) had too high error rates, even with ZNE being able to extrapolate meaningful expectation values that had a gradient (as opposed to the examples in Figure~\ref{fig:appendix_HOP_ZNE_converged_0.5}) often times the effective quantum volume test would fail. Examples of this are shown in Figure~\ref{fig:example_QV_not_passing_threshold} in the form of cumulative HOP plots. Note that in the top right hand plot the threshold is only barely not passed, whereas in the other plots the mean ZNE HOP distribution is more clearly not passed. This shows that, as with previous Quantum Volume studies that sample a range of device settings and qubits~\cite{QV_in_practice}, there is a distribution of performances, and the utilization of ZNE does not uniformly automatically make the measured results better. Near term quantum error mitigation does require tuning the of the experimental settings, and is still dependent on what errors are encountered on the device.

\clearpage

\setlength\bibitemsep{0pt}
\printbibliography